\documentclass[aps,pra,twocolumn,amsmath,amssymb,nofootinbib,showpacs,superscriptaddress]{revtex4-2}
\usepackage[english]{babel}
\usepackage{latexsym}
\usepackage{graphicx}
\usepackage{epstopdf}
\usepackage{epsfig}
\usepackage{color}
\usepackage{bm}
\usepackage{amsmath}
\usepackage{amssymb}
\usepackage{amsthm}
\usepackage{dcolumn}
\usepackage{float}
\usepackage{hyperref}
\usepackage{color}
\usepackage{cleveref}
\usepackage[svgnames,dvipsnames]{xcolor}
\usepackage{enumerate}
\usepackage{braket}

\hypersetup{hidelinks,colorlinks=true,allcolors=DarkBlue}

\DeclareMathOperator{\Tr}{Tr}
\DeclareMathOperator{\spec}{spec}

\theoremstyle{remark}
\newtheorem{remark}{Remark}

\renewcommand*{\Re}{\mathop{\mathrm{Re}}\nolimits}
\renewcommand*{\Im}{\mathop{\mathrm{Im}}\nolimits}

\begin{document} 

\title{Quantum master equations and steady states for the ultrastrong-coupling limit \\and the strong-decoherence limit}

\author{Anton Trushechkin}
\affiliation{Steklov Mathematical Institute of Russian Academy of Sciences, Moscow 119991, Russia}

\email{trushechkin@mi-ras.ru}

\date{\today}

\begin{abstract}
In the framework of theory of open quantum systems,
we derive quantum master equations for the ultrastrong system-bath coupling regime and, more generally, the strong-decoherence regime. In this regime, the strong decoherence is complemented by slow relaxation processes. We use a generalization of the F\"{o}rster and modified Redfield peturbation theories known in theory of excitation energy transfer. Also, we show that the mean force Gibbs state in the corresponding limits are stationary for the derived master equations.
\end{abstract}

\maketitle

\section{Introduction} 

The dynamics of quantum systems strongly coupled to the environment (bath) is an actively developing direction in theory of open quantum systems, which has many applications in physics, especially in quantum thermodynamics \cite{KatzKosloff,Nazir,Dou,Strasberg,Rivas}. Though the approximation of weak system-bath coupling is widely used  and many classical results of theory open quantum systems are obtain in the framework of this approximation (including celebrated Redfield and Davies quantum master equations \cite{Redfield, Davies,Davies2}), this approximation is too restrictive in many physical systems. 


If we cannot apply the weak-coupling approximation and the problem is not exactly solvable, then we have three possibilities. We can apply one of the numerically exact methods, such as the hierarchical equations of motion (HEOM) \cite{TaniKubo,IFl,Tani}, an approximation of an infinite bath by a finite number of oscillation modes with Markovian dynamics (see Refs.~\cite{Vega} for a review and, e.g., Refs.~\cite{Tamapre,Tama,GarrawayPetruc,TereFinT,TereSeveralBath} for recent results), etc. The second possibility is to map the system into a transformed system (which includes some degrees of freedom of the bath) for which the weak-coupling approximation can be used.  Examples are the collective coordinate method \cite{Strasberg,Lambert} and the polaron transformation approach \cite{Polaron0,Polaron}. The third possibility is to develop a perturbation theory different from the weak-coupling perturbation theory. Well-known examples are the singular-coupling limit \cite{Gorini,Palmer,AccFriLu} and the low-density limit \cite{Dumcke,Pechen,BP}.

The aim of this paper is to develop a perturbation theory for the ultrastrong-coupling limit, which is opposite to the weak-coupling limit. There is an increasing interest to such regime in the last years \cite{Kawai,CresserAnders,UstrongChargedQubits,LambertNori}. Moreover, we argue that the ultrastrong-coupling limit can be considered a particular case of a more general approximation called the strong-decoherence limit. 

We show that, in fact, the  so called F\"orser approximation from excitation energy transfer (EET) theory \cite{MK,Valkunas,YangFl,NovoGrond,Seibt} describes ultrastrong system-bath coupling. We generalize this approximation to the general setting of open quantum systems. More general strong-decoherence limit also involves a generalization of the modified Redfield theory also widely used in theory of EET.

The F\"orser theory \cite{Forster1,Forster2} is the basic theory of EET and is based on an assumption that the couplings between the local excitations are much weaker than the system-bath coupling (which has the form of local decoherence). So, this is the case of strong system-bath coupling. 

The modified Redfield approach \cite{mRedf} (in contrast to the weak-coupling, or the standard Redfield approach) treats the pure decoherence part of the system-bath interaction non-perturbatively. In other words, only the off-diagonal part of the system-bath interaction Hamiltonian (in the eigenbasis of the isolated system Hamiltonian) is assumed to be small. This theory is basic for understanding the coherent EET in biological light-harvesting complexes \cite{YangFl,NovoGrond,Seibt}.

We adapt these approximations, which have proved to be very useful in theory of EET, to the general framework of open quantum systems  and also to generalize and unify them. In particular, we allow for the collective action of strong pure decoherence on subspaces of the system's Hilbert space and weak-coupling dynamics inside these subspaces. For example, a subspace may correspond to degenerate or nearly degenerate energy levels. This generalization is important since the modified Redfield theory is known to fail in this case \cite{NovoGrond,NovoGrond2013,NovoGrond2017}. Moreover,  this is not a technical, but a fundamental limitation of the modified Redfield theory in its usual formulation (where the pure decoherence acts on different energy eigenstates separately, not collectively) \cite{TrushJCP}.

Another motivation of our work is to derive a steady state for the ultrastrong-coupling regime and the strong-decoherence regime. This is of a particular importance due to a discussion in papers~\cite{Kawai,CresserAnders} about the correct form of the steady state for the ultrastrong-coupling regime. We show that the steady state corresponds to the so called mean force Gibbs state, which confirms the conjecture of Ref.~\cite{CresserAnders}. We explain in which sense the state conjectured in Ref.~\cite{Kawai} can be considered stationary. Note that the mean force Gibbs state  differs from the usual Gibbs state with respect to the system Hamiltonian.

The following text is organized as follows. In Sec.~\ref{SecModel}, we introduce the model: an arbitrary system with a purely discrete spectrum interacting with the thermal bosonic bath. In Sec.~\ref{SecSimp}, we consider a simple particular case of the ultrastrong-coupling (or strong-decoherence) approximation. It generalized the F\"orster approximation and includes the non-degenerate ultrastrong interaction. We derive the corresponding master equation and its steady state. This master equation describes the dynamics of the populations (the diagonal elements of the density matrix) in the so called pointer basis. The coherences are small (due to the strong-decoherence regime), but also can be calculated. This is done in Sec.~\ref{SecCoh}. Using this, we derive corrections to the steady state conjectured in Ref.~\cite{CresserAnders} and derived in the preceding Sec.~\ref{SecSimp}. In the end of Sec.~\ref{SecCoh}, the range of validity of the described approximation is discussed. In Sec.~\ref{SecGen}, we introduce general form of the strong-decoherence approximation, derive the corresponding master equation and its steady state.

\section{Model}\label{SecModel}

Let us consider the Hamiltonian of an open quantum system

\begin{equation}\label{EqH}
H=H_S+H_B+H_I,
\end{equation}
where three terms in the right hand side are a free system Hamiltonian (corresponding to a Hilbert space $\mathcal H_S$), a free bath Hamiltonian (corresponding to a Hilbert space $\mathcal H_B$), and a system-bath interaction Hamiltonian, respectively. $H_S$ is assumed to have purely discrete spectrum. We consider the bath of free harmonic oscillators:
\begin{equation}
H_B=\int \omega(\xi)a(\xi)^\dag a(\xi)\,d\xi,
\end{equation}
where $\omega(\xi)\geq0$ is the frequency of the mode $\xi$ and $a(\xi)^\dag$ and $a(\xi)$ are the creation and annihilation operators for the mode $\xi$. Let the interaction Hamiltonian take the form
{}
\begin{equation}\label{EqHI}
H_I=\sum_{\alpha=1}^MA_\alpha\otimes B_\alpha,
\end{equation}
where $A_\alpha$ are Hermitian operators on $\mathcal H_S$ and
\begin{equation}
B_\alpha=\int \big[\overline{d_\alpha(\xi)}a(\xi)+d_\alpha(\xi)a(\xi)^\dag
\big]\,d\xi.
\end{equation}

We assume that the state of the bath is thermal with an inverse temperature $\beta$ [not to be confused with the subindex $\beta$, which will be used as the second subindex in the double summations (\ref{EqHI})]. The generalization to the case of several thermal baths with different temperatures is straightforward. Let us denote this state 
\begin{equation}
\rho_B=Z^{-1}e^{-\beta H_B},\qquad Z=\Tr e^{-\beta H_B}. 
\end{equation}
Note that, due to an infinite number of oscillation modes, strictly speaking, $Z$ is ill-defined and $\rho_B$ is not a genuine density operator. To deal with such bath state, we can either consider the limit of a large finite number of modes or treat $\rho_B$ in the ``generalized'' sense, as a functional on the algebra of canonical commutation relations according to the formula
\begin{equation*}
\Tr[\rho_Ba^\dag(\xi)a(\xi')]\equiv
\langle a^\dag(\xi)a(\xi')\rangle=n_{\rm BE}[\omega(\xi)]\delta(\xi-\xi'),
\end{equation*}
and the Gaussian property.
Here $n_{\rm BE}(\omega)=(e^{\beta\omega}-1)^{-1}$ is the Bose--Einstein distribution and $\langle\,\cdot\,\rangle$ denotes the expectation with respect to the thermal state. We can associate $\Tr O\rho_B$ with $\langle  O\rangle$ for an arbitrary bath observable $O$.

\section{Non-degenerate ultrastong coupling}
\label{SecSimp}

\subsection{Description of the approximation}\label{SecForst}

Let us assume that all $A_\alpha$ in (\ref{EqHI}) commute and have the following form:

\begin{equation}\label{EqAalpha}
A_\alpha=\sum_{n}\theta_{\alpha n}\ket n\bra n,
\end{equation}
where $\{\ket n\}$ is an eigenbasis for all $A_\alpha$. For simplicity, in this section, we additionally assume that:
\begin{enumerate}[(i)]
\item For each $n$, there exists $\alpha$ such that $\theta_{\alpha n}\neq0$ (i.e., there is no subspace which does not interact with the bath).

\item Only the zero eigenvalue of each $A_\alpha$ may be degenerate. In other words, if $\theta_{\alpha n}=\theta_{\alpha m}$, then $\theta_{\alpha n}=\theta_{\alpha m}=0$.
\end{enumerate}
If these two conditions are met, the basis $\ket n$ is uniquely defined.

Then, the system Hamiltonian $H_S$ can be expressed as
\begin{equation}\label{EqHSforst}
H_S=\sum_n \varepsilon_n\ket n\bra n+
\sum_{n\neq m}J_{nm}\ket n\bra m,
\end{equation}
where $\varepsilon_n$ are real and $J_{nm}^*=J_{mn}$. 

Let us assume that $J_{nm}$ can be treated as small  with respect to the system-bath coupling strength. Note that, in theory of weak-coupling limit, the interaction Hamiltonian $H_I$ is treated as a small perturbation. Here we consider the situation where 
\begin{equation}
V=\sum_{n\neq m}J_{nm}\ket n\bra m
\end{equation}
can be treated as a small perturbation, while the rest part of the total Hamiltonian (\ref{EqH}) (which includes $H_I$) is not small:
\begin{equation}
H_0=\sum_n\varepsilon_n\ket n\bra n+H_B+H_I.
\end{equation} 
Since $H_I$ can be arbitrarily large, this case includes the ultrastrong-coupling limit. We refer to this case as the ``strong coupling limit of decoherence type'' because the unperturbed dynamics corresponds to decoherence in the basis $\{\ket n\}$. 

In theory of weak-coupling limit, the relaxation occurs in the eigenbasis of $H_S$. In contrast, here, it occurs in the common eigenbasis of $A_\alpha$. In the context of EET theory, this is the local excitation basis (see Remark~\ref{RemForster} below), while, in the context of measurement theory, it is called the pointer basis \cite{Kawai,Decoherence}. 

Also note that interaction (\ref{EqAalpha}) without a small correction $V$ (i.e., the case of pure decoherence, when $\{\ket n\}$ is an eigenbasis of both all $A_\alpha$ and $H_S$) was considered in recent paper~\cite{TDdecoh}. 

\begin{remark}
Note that ``literal'' ultrastrong-coupling limit
\begin{equation}\label{EqUltrastrongLim}
H=H_S+H_B+\lambda^{-1}H_I,\quad\lambda\to0,
\end{equation}
does not lead to a good theory. In particular, in this limit, the Hamiltonian may be unbounded from below. The described perturbation theory with respect to $V$ is the right formalization of the ultrastrong-coupling regime, which does not produce pathologies.
\end{remark}

\begin{remark}
Though we treat $V$ as a small perturbation, actually, it is not assumed that $J_{nm}$ are smaller than $\varepsilon_n$. It is only assumed that $J_{nm}$ are much smaller than $\theta_{\alpha n}$, which ensures strong decoherence. A detailed analysis will be given in Sec.~\ref{SecDecoh}.
\end{remark}

\begin{remark}\label{RemForster}
In theory of EET, a state $\ket n$ corresponds to the excitation of the local site (molecule) $n$, $\varepsilon_n$ are local excitation energies, and $J_{nm}$ are the dipole couplings between the molecules.  Usually, it is assumed that each molecule is coupled to its own phonon bath. In this case $\theta_{\alpha n}=\delta_{\alpha n}$ (here $\delta$ is the Kronecker symbol) and 
\begin{equation}\label{Eqdd0}
d_{\alpha}(\xi)d_\beta(\xi)\equiv0
\end{equation}
for $\alpha\neq\beta$. Eq.~(\ref{Eqdd0}) means that each mode may interact with at most one site. A violation of Eq.~(\ref{Eqdd0}) corresponds to correlated baths, which are also considered in theory of EET \cite{ForsterNoneq}. Then,
the described approximation is known as the F\"orster approximation \cite{Forster1,Forster2,MK,Valkunas,YangFl}. Here we adapt it to a general context of open quantum systems and, in Sec.~\ref{SecGen}, allow a more general system-bath interaction.
\end{remark}

\subsection{Projection operator}\label{SecProjForst}

The complicated joint dynamics of the system and the bath can be reduced to a simplified quantum master equation for a finite number of ``slow'' degrees of freedom only if the other (``fast'') degrees of freedom quickly relax to a state depending on the slow degrees of freedom. This is typically formalized on the language of projection operators \cite{RH,BP,MK,Valkunas}. Let $\mathcal P$ be a projection operator acting on the space of joint system-bath trace-class operators $\mathcal S$, and we assume that the joint system-bath state $\rho(t)$ quickly relaxes to the subspace $\mathcal S_0=\mathcal {PS}$. This operator should agree with the decomposition of the Hamiltonian into a reference part and a small perturbation (\ref{EqDecomp0}). Namely, the subspace $\mathcal S_0$ should be invariant with respect to the ``fast'' unitary dynamics $e^{-it{\mathcal L_0}}$, where ${\mathcal L_0}=[H_0,\,\cdot\,]$.

Let us analyze the fast dynamics $e^{-it{\mathcal L_0}}$. Due to the strong decoherence, an off-diagonal part of $\rho$ (in the basis $\{\ket n\}$) vanishes. So, $\mathcal S_0=\sum_{n}\mathcal S^{(n)}$, where $\mathcal S^{(n)}=\Pi_n\mathcal S\Pi_n$, $\Pi_n=\ket n\bra n$.

Also we can see that the fast dynamics in the subspaces $\mathcal S^{(n)}$ are decoupled from each other:
\begin{equation}
H_0=\sum_{n}\Pi_n H_0^{(n)}\Pi_n,
\end{equation}
where 
\begin{equation}
H_0^{(n)}=\varepsilon_n+H_B+\sum_\alpha \theta_{\alpha n}B_\alpha.
\end{equation}
So, it suffices to describe the fast dynamics inside each subspace. Moreover, as we see, the dynamics inside each subspace is reduced to the bath dynamics: The dynamics of the system is trivial. Let us express the bath Hamiltonian for each subspace as a Hamiltonian of displaced harmonic oscillators:
\begin{equation}
H_0^{(n)}=\varepsilon_n-\delta\varepsilon_n
+H_B^{(n)},
\end{equation}
where
\begin{gather}
H^{(n)}_B=\int \omega(\xi)a_n(\xi)^\dag a_n(\xi)\,d\xi,
\label{EqHBn}
\\
a_n(\xi)=a(\xi)+\frac{d_n(\xi)}{\omega(\xi)},
\\
d_n(\xi)=\sum_{\alpha=1}^M
\theta_{\alpha n}d_\alpha(\xi),\\
\delta\varepsilon_n=\int\frac{|d_n(\xi)|^2}{\omega(\xi)}\,d\xi,
\label{EqReorg}
\end{gather}
$I_S$ is the identity operator for the system.
The quantity $\delta\varepsilon_n$ 
is called the reorganization energy \cite{MK,Valkunas}. Often, it is considered a parameter characterizing the system-bath coupling strength. 

Under the free dynamics, the bath state quickly thermalizes, hence, $\rho^{(n)}\in\mathcal S^{(n)}$ quickly relaxes to the state of the form $\rho_S^{(n)}\otimes\rho_B^{(n)}$, where $\rho_B^{(n)}=Z^{-1} e^{-\beta H_B^{(n)}}$ is the thermal state with respect to the Hamiltonian $H_B^{(n)}$ of the displaced oscillators. Also, $\rho_S=\Tr_B\rho$ is the reduced system operator and $\rho_S^{(n)}=\Pi_n\rho_S\Pi_n$.

\begin{remark}
The conjecture of the thermalization of the bath under the free dynamics is also used in the weak coupling theory. Namely, the Born approximation states that the joint system bath state is always close to the product state $\rho_S(t)\otimes\rho_B$. The corresponding projection operator used in the weak-coupling theory is 
$\mathcal P\rho=(\Tr_B\rho)\otimes\rho_B$. This is a formalization of the assumption that the free dynamics quickly turn a bath state into the thermal state $\rho_B$. Here we have exactly the same assumption, with the only difference of an $n$-dependent displacement. Of course, such thermalization under unitary dynamics can take place only in the weak sense (i.e., in terms of averages of local and quasi-local observables), see rigorous results in Refs.~\cite{Bach,Frohlich}.
\end{remark}

Thus, we can define the projection operator $\mathcal P$ as follows:
\begin{eqnarray}
\mathcal P\rho&=&
\sum_{n}\Pi_n(\Tr_B\rho)\Pi_n\otimes\rho_B^{(n)}
\label{EqP}
\\
&=&\sum_{n}p_n\ket n\bra n\otimes\rho_B^{(n)},
\label{EqPforst}
\end{eqnarray}
where $p_n=\braket{n|\Tr_B\rho|n}$.

\begin{figure}[h]
\begin{centering}
\includegraphics[scale=1.3]{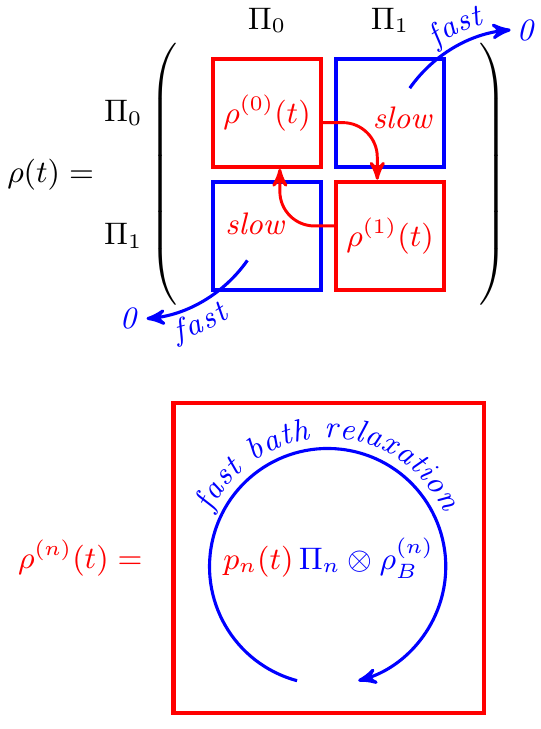}
\vskip -2mm
\caption
{\small
Fast and slow processes at the non-degenerate ultrastrong coupling.  For simplicity, here, the system is two-dimensional. The blue color denotes the fast processes and the red color denotes the slow processes and degrees of freedom, which are described by a quantum master equation. The notations are as in the text: $\Pi_n=\ket n\bra n$, $\rho_B^{(n)}$ are displaced bath thermal equilibrium sates, and $p_n(t)$ are populations of the states $\ket n$. The bottom part of the figure depicts the red squares [corresponding to $\rho^{(n)}(t)=\Pi_n\rho(t)\Pi_n$] in more detail. The fast processes are the decoherence (reduction to zero of the off-diagonal blocks) and the bath relaxation, while the slow process is the relaxation of the populations.}
\label{FigScheme}
\end{centering}
\end{figure}

The slow dynamics consists of transitions between different subspaces $\mathcal S^{(n)}$ governed by the perturbation $V$. Fast and slow dynamics are schematically represented in Fig.~\ref{FigScheme}.

Note that somewhat similar types of the projection operators and the corresponding master equations were considered in Refs.~\cite{EspositoGaspard,Budini,Breuer2006,Breuer2007,StrasbergHier}. In our approach, the  the bath equilibrium state depends on the system subspace. In these works, the bath (rather than the system) is decomposed into a sum of subspaces which determine different dynamics for the system.

\subsection{Master equation}

Denote $\mathcal L=[H,\,\cdot\,]$, $\mathcal L_0=[H_0,\,\cdot\,]$, and $L'=[V,\,\cdot\,]$. We will work in the interaction picture with respect to $H_0$ and denote $\rho(t)=e^{i\mathcal L_0 t}e^{-i\mathcal Lt}\rho_0$, 
\begin{eqnarray}
V(t)&=&e^{i\mathcal L_0t}V\nonumber\\
&=&\sum_{n\neq m}J_{nm}e^{i(\bar\varepsilon_n-\bar\varepsilon_m)t}
e^{iH_B^{(n)}t}\ket n\bra me^{-iH_B^{(m)}t},\qquad\label{EqVt}
\end{eqnarray} 
and $\mathcal L'(t)=[V(t),\,\cdot\,]=e^{i\mathcal L_0t}\mathcal L'e^{-i\mathcal L_0t}$. Here $$\bar\varepsilon_n=\varepsilon_n-\delta\varepsilon_n.$$ 

We assume that $\rho(0)=\mathcal P\rho(0)$. Since $\mathcal P\mathcal L'(t)\mathcal P=0$, the standard derivation of the Markovian second-order master equation \cite{BP,MK,RH} (see also Appendix~\ref{AppPQ}) leads to
\begin{equation}\label{EqMaster}
\mathcal P\dot\rho(t)=
\int_0^\infty d\tau\,
\mathcal P\mathcal L'(t)\mathcal L'(t-\tau)\mathcal P\rho(t).
\end{equation}
Since, in our case, 
\begin{equation}
\mathcal P\mathcal L_0=\mathcal L_0\mathcal P=0,
\end{equation}
Eq.~(\ref{EqMaster}) can be rewritten as
\begin{equation}\label{EqMasterAut}
\mathcal P\dot\rho(t)=
\int_0^\infty d\tau\,
\mathcal P\mathcal L'(\tau)\mathcal L'\mathcal P\rho(t).
\end{equation}

Substituting expression (\ref{EqPforst}) for $\mathcal P\rho$, we obtain
\begin{equation}\label{EqMasterForst}
\dot p_n(t)=\sum_{m\neq n}[\gamma_{nm}p_m(t)-\gamma_{mn}p_n(t)],
\end{equation}
where
\begin{equation}\label{Eqk}
\begin{split}
\gamma_{nm}&=2|J_{nm}|^2\Re\int_0^\infty d\tau\,
e^{i(\bar\varepsilon_m-\bar\varepsilon_n)\tau}
\langle e^{iH_B^{(m)}\tau}e^{-iH_B^{(n)}\tau}\rangle_m
\\
&=|J_{nm}|^2\int_{-\infty}^{+\infty} d\tau\,
e^{i(\bar\varepsilon_m-\bar\varepsilon_n)\tau}
\langle e^{iH_B^{(m)}\tau}e^{-iH_B^{(n)}\tau}\rangle_m
\end{split}
\end{equation}
and $\langle O\rangle_m=\Tr_B O\rho_B^{(m)}$.
So, since the off-diagonal part of $\rho_S(t)=\Tr_B\rho(t)$ quickly vanishes, the master equations describes only the diagonal elements $p_n(t)$.

\subsection{Mean-force Gibbs steady state}\label{SecStForst}
An explicit expression of $\gamma_{mn}$ will be obtained in the next subsection, but, already, a general expression (\ref{Eqk}) allows us to obtain a steady state. Let us change the variable of integration in Eq.~(\ref{Eqk}) by $s=-\tau-i\beta$:
\begin{eqnarray}
\gamma_{nm}
&&=|J_{nm}|^2\int_{-\infty}^{+\infty} ds\,
e^{-i(\bar\varepsilon_m-\bar\varepsilon_n)(s+i\beta)}
\nonumber
\\&&\qquad\times
\langle e^{-iH_B^{(m)}(s+i\beta)}e^{iH_B^{(n)}(s+i\beta)}\rangle_m
\nonumber
\\
&&=e^{\beta(\bar\varepsilon_m-\bar\varepsilon_n)}
|J_{nm}|^2
\int_{-\infty}^{+\infty} ds\,
e^{i(\bar\varepsilon_n-\bar\varepsilon_m)s}
\nonumber
\\
&&\qquad\times
\langle e^{-iH_B^{(m)}s}e^{iH_B^{(n)}s}
e^{-\beta H_B^{(n)}}e^{\beta H_B^{(m)}}
\rangle_m
\nonumber
\\
&&=e^{\beta(\bar\varepsilon_m-\bar\varepsilon_n)}\gamma_{mn},
\label{EqDetBalForst}
\end{eqnarray}
where we have used
\begin{equation}\label{EqOmn}
\langle Oe^{-\beta H_B^{(n)}}e^{\beta H_B^{(m)}}\rangle_m=
\langle O\rangle_n.
\end{equation}

We have obtained the detailed balance conditions. Then, the following populations and the corresponding density operator are stationary:
\begin{equation}\label{Eqpst}
p^{(\rm st)}_n=Z_S^{-1}e^{-\beta\bar\varepsilon_n},
\end{equation}
where 
\begin{equation}
Z_S=\sum_m e^{-\beta\bar\varepsilon_n}
=\Tr e^{-\beta\sum_n\Pi_n \bar H_S\Pi_n},
\end{equation}
or, equivalently,
\begin{equation}\label{EqStForst}
\rho_S^{(\rm st)}=Z_S^{-1}e^{-\beta\sum_n\Pi_n \bar H_S^{\rm (d)}\Pi_n}.
\end{equation}
Here
\begin{equation}
\bar H_S^{\rm (d)}=\sum_n\bar\varepsilon_n\ket n\bra n
\end{equation}
is the ``renormalized'' diagonal part of the system Hamiltonian. 

It is expected that, if the system is ergodic, i.e., there is no non-trivial proper subspace of the system Hilbert space $\mathcal H_S$, which is invariant with respect to the system-bath dynamics, then the reduced state of the system  converges to the so called mean-force Gibbs state 
\begin{equation}\label{EqMFG}
\rho_S^{(\rm MFG)}=\Tr_B Z^{-1}e^{-\beta H}, \qquad Z=\Tr e^{-\beta H}. 
\end{equation}
State~(\ref{EqStForst}), obviously, coincides with the mean-force Gibbs state in the limit (small $J_{nm}$). Also, it coincides with the mean-force Gibbs state calculated for the ``literal'' ultrastrong-coupling limit (\ref{EqUltrastrongLim}) \cite{CresserAnders}. 

\begin{remark}\label{RemCounter}
Note that, in the ``literal'' ultrastrong-coupling limit, the reorganization energy (\ref{EqReorg}) tends to infinity and we should artificially introduce the corresponding counter-term in the Hamiltonian. In the presented strong-decoherence limit, we are free of such divergence and there is no need for its introduction. A small difference with the result of Ref.~\cite{CresserAnders} is caused by this difference: We have not introduced this counter-term to the original Hamiltonian (\ref{EqH}). We could do so to obtain exactly the same result.
\end{remark}

If the system is ergodic, i.e., in our case, there is no non-trivial proper subspace of states $\{\ket n\}$ isolated form the other states, than stationary state~(\ref{EqStForst}) is unique.

Not that a mathematically rigorous proof of convergence to the state (\ref{EqStForst}) for the spin-boson model and super-Ohmic spectral densities [see Eqs.~(\ref{EqJab}) and~(\ref{EqDL}) below] such that $\mathcal J(\omega)=O(\omega^3)$ as $\omega\to0$ has been  given in Refs.~\cite{MerkliNesterovDimer,MerkliDimer}. Here, we give a ``physically rigorous'' proof (i.e., not mathematically rigorous, but, nevertheless, based on the microscopic model and physically plausible assumptions) under the condition that the spectral densities~(\ref{EqJab}) are either super-Ohmic or Ohmic. A sub-Ohmic spectral densities lead to divergent bath correlation functions~(\ref{EqCJab}) and reorganization energies $\delta\varepsilon_{\alpha\beta}$.

\subsection{Rate constants}

In order to calculate the rate constants $\gamma_{nm}$, we should obtain an explicit expression for $\langle e^{iH_B^{(m)}t}e^{-iH_B^{(n)}t}\rangle_m$. It turns out that (see Appendix~\ref{AppCalc})
\begin{multline}\label{EqIntegrand}
\langle e^{iH_B^{(m)}t}e^{-iH_B^{(n)}t}\rangle_m
\equiv\zeta_{nm}(t)
\\
=\exp\Big[
-\sum_{\alpha\beta}
(\theta_{\alpha n}-\theta_{\alpha m})
(\theta_{\beta n}-\theta_{\beta m})
(g_{\alpha\beta}(t)+it
\delta\varepsilon_{\alpha\beta})
\Big], 
\end{multline}
where
\begin{equation}
\delta\varepsilon_{\alpha\beta}=
\int\frac{\overline{d_\alpha(\xi)}d_\beta(\xi)}{\omega(\xi)}\,d\xi
\end{equation}
and
\begin{equation}
g_{\alpha\beta}(t)=\int_0^t ds_1\int_0^{s_1}ds_2\,
C_{\alpha\beta}(s_2).
\end{equation}
Here
\begin{equation}
C_{\alpha\beta}(t)=
\langle e^{iH_Bt}B_\alpha e^{-iH_Bt}B_\beta\rangle
\end{equation}
are the bath correlation functions. Recall that $\langle O\rangle=\Tr_B O\rho_B$ and $\rho_B=Z^{-1}e^{-\beta H_B}$. In spectroscopy, $g_{\alpha\beta}(t)$ are called the lineshape functions since the absorption and fluorescence spectra are expressed through them. We assume that the bath correlation functions $C_{\alpha\beta}(t)$ are integrable. Then the functions $g_{\alpha\beta}(t)$ grow linearly with $t$ for large $t$.

It is worthwhile to note that $\gamma_{nm}\to0$ in both the considered limit of small $J_{nm}$ and the ``literal'' ultrastrong-coupling limit $\theta_{\alpha n}\to\infty$ (for all $\alpha$ and $n$), which was expected in view of the quantum Zeno effect. Slow dynamics of populations is a correction to the quantum Zeno effect. 

Note also that 
\begin{subequations}
\begin{eqnarray}
C_{\alpha\beta}^*(t)&=&C_{\beta\alpha}(-t), \\
g_{\alpha\beta}^*(t)&=&g_{\beta\alpha}(-t),
\label{Eqgstar}\\ 
\zeta_{nm}^*(t)&=&\zeta_{nm}(-t),
\end{eqnarray}
\end{subequations}
\begin{equation}
\begin{split}
C_{\alpha\beta}(t)=\int d\xi\,
\Big\{
&\overline{d_\alpha(\xi)}d_\beta(\xi)
\big(n_{\rm BE}[\omega(\xi)]+1\big)e^{-i\omega(\xi)t}
\\
+
&d_\alpha(\xi)\overline{d_\beta(\xi)}
n_{\rm BE}[\omega(\xi)]e^{i\omega(\xi)t}
\Big\}.
\end{split}
\end{equation}
If we introduce the spectral densities
\begin{equation}\label{EqJab}
\mathcal J_{\alpha\beta}(\omega)
=\int 
\overline{d_\alpha(\xi)}d_\beta(\xi)
\delta[\omega(\xi)-\omega]\,d\xi,
\end{equation}
then the bath correlation functions and the lineshape functions can be expressed as
\begin{equation}\label{EqCJab}
\begin{split}
C_{\alpha\beta}(t)=\int_0^\infty d\omega\,
\Big\{
&\mathcal J_{\alpha\beta}(\omega)
[n_{\rm BE}(\omega)+1]e^{-i\omega t}
\\
+
&\mathcal J_{\beta\alpha}(\omega)
n_{\rm BE}(\omega)e^{i\omega t}
\Big\}
\end{split}
\end{equation}
and
\begin{equation}\label{EqgJab}
\begin{split}
g_{\alpha\beta}(t)=\!-\!\int_0^\infty\!d\omega
\Big\{
&\frac{\mathcal J_{\alpha\beta}(\omega)}{\omega^2}
[n_{\rm BE}(\omega)+1]
(e^{-i\omega t}+i\omega t-1)
\\
+
&\frac{\mathcal J_{\beta\alpha}(\omega)}{\omega^2}
n_{\rm BE}(\omega)
(e^{i\omega t}-i\omega t-1)
\Big\}.
\end{split}
\end{equation}
We assume that the spectral functions are either Ohmic or super-Ohmic [i.e., $O(\omega)$ as $\omega\to0$] and integrable. In this case, all the integrals converge.

\subsection{Example: Spin-boson model at ultrastrong coupling}
\label{SecExamp}

Let us consider the spin-boson model as an example. Let the system Hamiltonian be
\begin{equation}
H_S=\varepsilon\sigma_z=\varepsilon(\ket1\bra1-\ket0\bra0).
\end{equation}
Let the system ultrastrongly interact with a single thermal bath with the inverse temperature $\beta$ and the interaction Hamiltonian 
\begin{equation}
H_I=\sigma_x\otimes B=(\ket+\bra+-\ket-\bra-)\otimes B,
\end{equation}
where $\ket{\pm}=(\ket1\pm\ket0)/\sqrt2$, $\sigma_z$ and $\sigma_x$ are the Pauli matrices and
\begin{equation}
B=\int \Big[\overline{d(\xi)}a(\xi)+d(\xi)a(\xi)^\dag
\Big]\,d\xi
\end{equation}
[i.e., the sum (\ref{EqHI}) contains only one term and the subindices $\alpha$ and $\beta$ disappear everywhere]. 

In the weak coupling regime, relaxation occurs in the eigenbasis $\{\ket0,\ket1\}$ of $\sigma_z$. In the ultrastrong-coupling regime, it occurs in the eigenbasis $\{\ket+,\ket-\}$ of 
$\sigma_x$. So, $n\in\{+,-\}$, $\theta_\pm=\pm1$. Let us express the $H_S$ Hamiltonian in this basis:
\begin{equation}
H_S=\varepsilon(\ket+\bra-
+\ket-\bra+),
\end{equation}
hence $\varepsilon_+=\varepsilon_-=0$ and $J_{-+}=J_{+-}=\varepsilon$.
The application of the general formula (\ref{EqIntegrand}) gives
\begin{equation*}
\langle e^{iH_B^{(+)}t}e^{-iH_B^{(-)}t}\rangle_+
=
\langle e^{iH_B^{(-)}t}e^{-iH_B^{(+)}t}\rangle_-
=
e^{-4[g(t)+it\delta\varepsilon]}.
\end{equation*}

For the simulation, we choose the Drude--Lorentz spectral density:
\begin{equation}\label{EqDL}
\mathcal J(\omega)\equiv\int
|d(\xi)|^2
\delta(\omega(\xi)-\omega)\,dk=
\frac{2\eta\Omega\omega}{\pi(\omega^2+\Omega^2)}
\end{equation}
with $\eta=100~\rm{cm}^{-1}$ and $\Omega^{-1}=100~\rm{fs}$ ($\Omega\approx53.08~\rm{cm}^{-1}$). The temperatures of the bath is $T=300~\rm{K}$. The parameter $\varepsilon$ is $\varepsilon=10~\rm{cm}^{-1}$. The reorganization energy can be calculated as
\begin{equation}
\delta\varepsilon=\int_0^\infty\frac{\mathcal J(\omega)}\omega
\,d\omega=\eta.
\end{equation}

The bath correlation function can be expressed as
\begin{multline}\label{EqCJ}
C(t)=\langle e^{iH_Bt}Be^{-iH_Bt}B\rangle
\\
=
\int_0^\infty\mathcal J(\omega)
\left[
\coth\left(\frac{\beta\omega}2\right)\cos\omega t-i\sin\omega t
\right]d\omega.
\end{multline}

We adopt the high-temperature approximation $\beta\Omega\ll1$. For example, for the used value $\Omega\approx53.08~\rm{cm}^{-1}$ and the temperature $T=300$~K, we have $\beta\Omega\approx0.24$. We have used that $\beta=1/k_{\rm B}T$, where $k_{\rm B}\approx0.734~\rm{cm^{-1}/K}$ is the Boltzmann constant. In this case, $\coth(\beta\omega/2)$ in Eq.~(\ref{EqCJ}) can be approximated as $2/(\beta\omega)$ and
\begin{eqnarray}
C(t)&\approx&\eta\Omega
\left(\frac{2}{\beta\Omega}-i\right)e^{-\Omega t},\\
g(t)&\approx&
\frac{\eta}{\Omega}
\left(\frac{2}{\beta\Omega}-i\right)
(e^{-\Omega t}+\Omega t-1).
\label{EqgDL}
\end{eqnarray}

\begin{figure}[h]
\begin{centering}
\includegraphics[width=\linewidth]{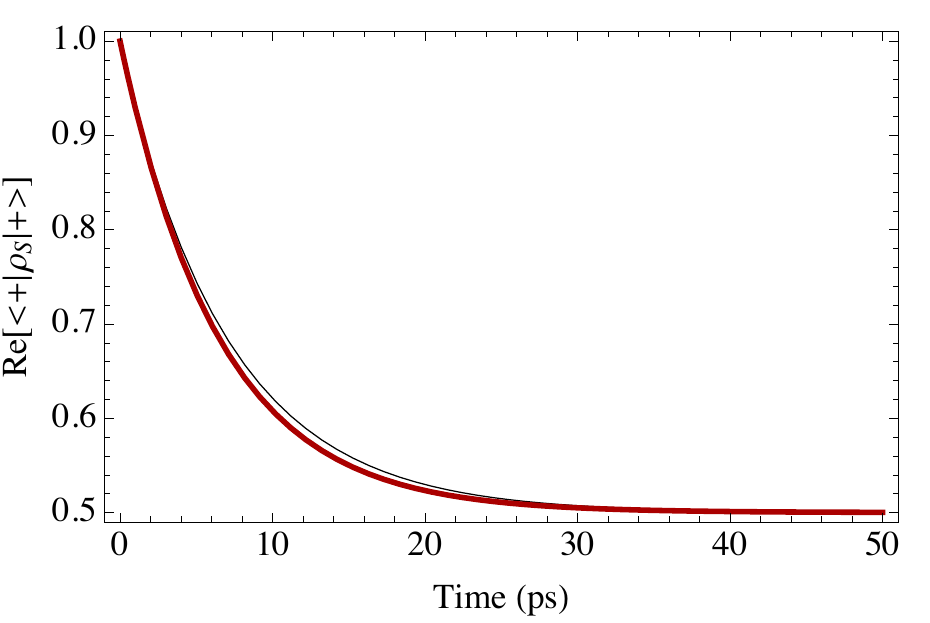}
\vskip -2mm
\caption
{\small
Spin-boson model at ultrastrong coupling: performance of the master equation (\ref{EqMasterForst}) (thick red line) in comparison with the numerically exact (but much more computationally expensive) method of HEOM (thin black line).
}
\label{FigSpinBoson}
\end{centering}
\end{figure}

For our parameters, $\gamma_{+-}=\gamma_{-+}\approx0.41~\text{cm}^{-1}$ or $0.077~\text{ps}^{-1}$, so, the characteristic relaxation time is $(2\gamma_{+-})^{-1}\approx6.5~{\rm ps}$. In Fig.~\ref{FigSpinBoson}, we compare the solution using master equation (\ref{EqMasterForst}) with the numerically exact (but much more computationally expensive) method of hierarchical equations of motion (HEOM) in the high-temperature approximation \cite{IFl}. The initial state is 
\begin{equation}\label{EqIni}
\rho(0)=\ket+\bra+\otimes\rho_B.
\end{equation} 
We see almost ideal agreement.

\section{Dynamics of coherences}\label{SecCoh}

\subsection{Corrections to the diagonal steady state}

We have  determined the populations, i.e., the diagonal matrix elements of the system density operator in the pointer basis $\{\ket n\}$. The projection operator $\mathcal P$ projects onto the diagonal states. Hence, to obtain the off-diagonal elements (coherences), one needs the part $\mathcal Q\rho(t)$, where $\mathcal Q=1-\mathcal P$  \cite{TrushJCP}. If we denote 
\begin{equation}
\rho_{nm}(t)=\braket{n|\Tr_B\{e^{-iH_0t}\rho(t)e^{iH_0t}\}|m},
\end{equation}
$n\neq m$
(i.e., the coherences in the Schr\"odinger picture), then, assuming again $\rho(0)=\mathcal P\rho(0)$ and using Eq.~(\ref{EqQ1Pre}) for $\mathcal Q\rho(t)$, we obtain
\begin{gather}
\rho_{nm}(t)=
\braket{n|\Tr_B\{e^{-iH_0t}[\mathcal Q\rho(t)]e^{iH_0t}\}|m}
\label{EqCoh}
\\
=-i
\int_0^td\tau
\braket{n|
\Tr_B
\{
e^{-iH_0t}
\mathcal L'(t-\tau)\mathcal P\rho(t-\tau)
e^{iH_0t}
\}
|m}
\nonumber
\end{gather}
Using $[\mathcal P\rho(t),H_0]=0$ and Eq.~(\ref{EqVt}), we obtain
\begin{equation*}
\begin{split}
&\rho_{nm}(t)=
-i
\int_0^td\tau
\braket{n|
\Tr_B
\{
\mathcal L'(-\tau)\mathcal P\rho(t-\tau)
\}
|m}
\\
&=iJ_{nm}\int_0^td\tau\,
p_n(t-\tau)
e^{i(\bar\varepsilon_m-\bar\varepsilon_n)\tau}
\langle e^{iH_B^{(m)}\tau}e^{-iH_B^{(n)}\tau}\rangle_n
\\
&-iJ_{nm}\int_0^td\tau\,
p_m(t-\tau)
e^{i(\bar\varepsilon_m-\bar\varepsilon_n)\tau}
\langle e^{iH_B^{(m)}\tau}e^{-iH_B^{(n)}\tau}\rangle_m.
\end{split}
\end{equation*}
Here, the integrands have been already calculated in Eq.~(\ref{EqIntegrand}). We can again [like in the derivation of Eq.~(\ref{EqMaster})] apply the Markovian approximation and replace $p_n(t-\tau)$ and $p_m(t-\tau)$ by $p_n(t)$ and $p_m(t)$ since  they evolve on larger time scales than the decay rate of the integrands. Thus, finally, we obtain
\begin{equation}\label{EqCohEq}
\begin{split}
\rho_{nm}(t)&=
iJ_{nm}p_n(t)\int_0^t\zeta_{mn}^*(\tau)
\,e^{i(\bar\varepsilon_m-\bar\varepsilon_n)\tau}
d\tau
\\
&-iJ_{nm}p_m(t)\int_0^t\zeta_{nm}(\tau)
\,e^{i(\bar\varepsilon_m-\bar\varepsilon_n)\tau}
d\tau.
\end{split}
\end{equation}

For large times, we can extend the upper limit of integration in Eq.~(\ref{EqCohEq}) to infinity and obtain constant coefficients after the populations. So, for large times, the dynamics of the coherences is driven by the populations. If we take the limit $t\to\infty$ and substitute the time-dependent populations $p_n(t)$ by their stationary values (\ref{Eqpst}), then we will obtain the steady-state coherences as the first-order corrections to the diagonal steady state (\ref{EqStForst}):
\begin{equation}
\begin{split}
\rho^{(\rm st)}_{nm}&=
iJ_{nm}p^{(\rm st)}_n\int_0^\infty\zeta_{mn}^*(\tau)
\,e^{i(\bar\varepsilon_m-\bar\varepsilon_n)\tau}
d\tau
\\
&-iJ_{nm}p^{(\rm st)}_m\int_0^\infty\zeta_{nm}(\tau)
\,e^{i(\bar\varepsilon_m-\bar\varepsilon_n)\tau}
d\tau.
\end{split}
\end{equation}
Since
\begin{multline*}
\Re\int_0^\infty\zeta_{nm}(\tau)
\,e^{i(\bar\varepsilon_m-\bar\varepsilon_n)\tau}\,d\tau
\\=
\frac12
\int_{-\infty}^{+\infty}\zeta_{nm}(\tau)
\,e^{i(\bar\varepsilon_m-\bar\varepsilon_n)\tau}\,d\tau,
\end{multline*}
then, due to Eq.~(\ref{EqDetBalForst}) and (\ref{Eqpst}), 
\begin{multline*}
p^{(\rm st)}_m\Re\int_0^\infty\zeta_{nm}(\tau)
\,e^{i(\bar\varepsilon_m-\bar\varepsilon_n)\tau}\,d\tau
\\
=
p^{(\rm st)}_n\Re\int_0^\infty\zeta_{mn}(\tau)
\,e^{i(\bar\varepsilon_n-\bar\varepsilon_m)\tau}\,d\tau.
\end{multline*}
Hence,
$\Im\rho^{(\rm st)}_{nm}=0$ and
\begin{equation}
\begin{split}
\rho^{(\rm st)}_{nm}&=
J_{nm}p^{(\rm st)}_n\Im\int_0^\infty\zeta_{mn}(\tau)
\,e^{i(\bar\varepsilon_n-\bar\varepsilon_m)\tau}
d\tau
\\
&+J_{nm}p^{(\rm st)}_m\Im\int_0^\infty\zeta_{nm}(\tau)
\,e^{i(\bar\varepsilon_m-\bar\varepsilon_n)\tau}
d\tau.
\end{split}
\end{equation}

\begin{figure}[h]
\begin{centering}
\includegraphics[width=\linewidth]{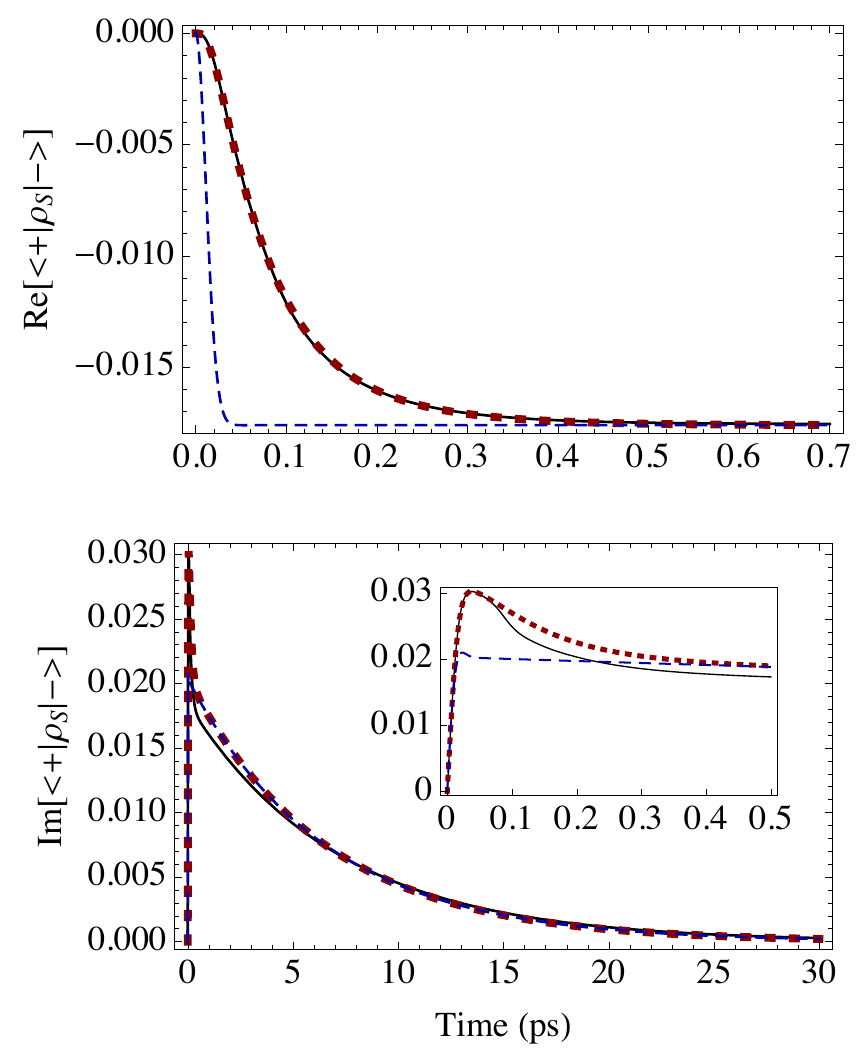}
\vskip -2mm
\caption
{\small
Dynamics of the coherence for the 
 spin-boson model
at ultrastrong coupling: formula (\ref{EqCohEq}) (blue dashed line), formula (\ref{EqCohSum}) with non-equilibrium short-term corrections (thick red dashed line) in comparison with HEOM (solid black line).
}
\label{FigSpinBosonCoh}
\end{centering}
\end{figure}

Consider again our example from Sec.~\ref{SecExamp}. For the same parameters and the same initial state~(\ref{EqIni}), we have calculated the coherence $\braket{+|\rho_S(t)|-}$ using formula~(\ref{EqCohEq}) and compared it with the numerically exact result by HEOM, see Fig.~\ref{FigSpinBosonCoh}. We see that formula~(\ref{EqCohEq}) correctly predicts the dynamics of the coherence on large times and, in particular, the steady-state coherence. 

However, on initial short times, it gives significant error because the assumption $\rho(0)=\mathcal P\rho(0)$ is not satisfied for state (\ref{EqIni}). We call a state $\rho$ equilibrium if $\rho=\mathcal P\rho$: the bath state is in the equilibrium, which depends on the system state $\ket n$. In this sense, state~(\ref{EqIni}) is nonequilibrium. In the next subsection, we derive the nonequilibrium corrections for  initial short times.

\subsection{Non-equilibrium corrections}

Consider now the initial system-bath state of the form
\begin{equation}
\rho(0)=\sum_n p_n(0)\ket n\bra n\otimes\rho_B.
\end{equation}
Thus, now 
\begin{equation}
\mathcal Q\rho(0)=\sum_n p_n(0)\ket n\bra n\otimes(\rho_B-\rho_B^{(n)}).
\end{equation}
The substitution of the expression Eq.~(\ref{EqQ1Inhom}) for $\mathcal Q\rho(t)$ into Eq.~(\ref{EqCoh}) gives
\begin{equation}\label{EqCohSum}
\rho_{nm}(t)=\rho_{nm}^{(\rm eq)}(t)+\rho_{nm}^{(\rm noneq)}(t),
\end{equation}
where $\rho_{nm}^{(\rm eq)}(t)$ is given by Eq.~(\ref{EqCohEq}) and
\begin{equation}\label{EqCohNoneq}
\begin{split}
\rho_{nm}^{(\rm noneq)}(t)=
i&J_{nm}\int_0^td\tau\,
e^{i(\varepsilon_m-\varepsilon_n)\tau}
\\
\times
\big\{
&p_n(0)[\zeta_{nmn}^*(t,\tau)
-
e^{-i(\delta\varepsilon_m-\delta\varepsilon_n)\tau}
\zeta^*_{mn}(\tau)]
\\
-&p_m(0)
[\zeta_{mnm}(t,\tau)-
e^{-i(\delta\varepsilon_m-\delta\varepsilon_n)\tau}
\zeta_{nm}(\tau)]
\big\}.
\end{split}
\end{equation}
Here 
\begin{equation*}
\begin{split}
\zeta_{mnl}(t,\tau)=
\Big\langle
&\exp\Big\{i\Big(H_B+\sum_\alpha\theta_{\alpha m}B_\alpha\Big)t\Big\}
\\
\times&\exp\Big\{-i\Big(H_B+\sum_\alpha\theta_{\alpha n}
B_\alpha\Big)\tau\Big\}
\\
\times&
\exp\Big\{-i\Big(
H_B+\sum_\alpha\theta_{\alpha l}B_\alpha
\Big)(t-\tau)\Big\}
\Big\rangle
\end{split}
\end{equation*}
In Appendix~\ref{AppCalc}, we show that
\begin{equation}\label{EqZetaNoneq}
\begin{split}
&\zeta_{mnl}(t,\tau)=
\exp\biggl\{
-\sum_{\alpha\beta}
(\theta_{\alpha n}-\theta_{\alpha m})
(\theta_{\beta n}-\theta_{\beta l})g_{\alpha\beta}(\tau)
\\
&+
\sum_{\alpha\beta}
(\theta_{\alpha n}-\theta_{\alpha m})
\big[\theta_{\beta m}g^*_{\alpha\beta}(t)-\theta_{\beta l}g_{\alpha\beta}(t)\big]
\\
&+
\sum_{\alpha\beta}
(\theta_{\alpha n}-\theta_{\alpha l})
\big[\theta_{\beta l}g_{\alpha\beta}(t-\tau)-
\theta_{\beta m}g^*_{\alpha\beta}(t-\tau)\big]
\bigg\}.
\end{split}
\end{equation}
Note that analogous traces for fermionic baths were calculated in Ref.~\cite{TDdecoh}. Thus,
\begin{equation}
\begin{split}
&\zeta_{mnm}(t,\tau)=
\exp\biggl\{
-\sum_{\alpha\beta}
(\theta_{\alpha n}-\theta_{\alpha m})
(\theta_{\beta n}-\theta_{\beta m})g_{\alpha\beta}(\tau)
\\
&-
2i\sum_{\alpha\beta}
(\theta_{\alpha n}-\theta_{\alpha m})
\theta_{\beta m}
\Im[g_{\alpha\beta}(t)-g_{\alpha\beta}(t-\tau)]
\bigg\}.
\end{split}
\end{equation}

Let us show that 
\begin{equation}\label{EqZetaLim}
\lim_{t\to\infty}\zeta_{mnm}(t,\tau)=
e^{-i(\delta\varepsilon_m-\delta\varepsilon_n)\tau}
\zeta_{nm}(\tau)
\end{equation}
and, thus, nonequilibrium correction (\ref{EqCohNoneq}) vanishes for long times. Since both $\zeta_{mnm}(t,\tau)$ and $\zeta_{nm}(\tau)$ vanish for large $\tau$, it is sufficient to prove that
\begin{equation}\label{EqImLim}
-2\lim_{t\to\infty}
\Im[g_{\alpha\beta}(t)-g_{\alpha\beta}(t-\tau)]
=(\delta\varepsilon_{\alpha\beta}+\delta\varepsilon_{\beta\alpha})\tau
\end{equation}
for an arbitrary constant $\tau$. From Eqs.~(\ref{Eqgstar}) and (\ref{EqgJab}), we have
\begin{equation}\label{EqImg}
\begin{split}
-2\Im g_{\alpha\beta}(t)=i[g_{\alpha\beta}(t)&-g_{\beta\alpha}(-t)]
\\
=-i
\int_0^\infty d\omega\,
\Big\{
&\frac{\mathcal J_{\alpha\beta}(\omega)}{\omega^2}
(e^{-i\omega t}+i\omega t-1)
\\
-
&\frac{\mathcal J_{\beta\alpha}(\omega)}{\omega^2}
(e^{i\omega t}-i\omega t-1)
\Big\}
\end{split}
\end{equation}
and
\begin{multline}
-2\Im[g_{\alpha\beta}(t)-g_{\alpha\beta}(t-\tau)]
=\tau\!\int_0^\infty\!d\omega\,
\frac{\mathcal J_{\alpha\beta}(\omega)
+\mathcal J_{\beta\alpha}(\omega)}{\omega}
\\
+\int_0^\infty d\omega\,
\Big\{
\frac{\mathcal J_{\alpha\beta}(\omega)}{\omega^2}
e^{-i\omega t}(1-e^{i\omega \tau})
\\
-\frac{\mathcal J_{\beta\alpha}(\omega)}{\omega^2}
e^{i\omega t}
(1-e^{-i\omega \tau})
\Big\}.
\end{multline}
Here, the first integral is exactly the right-hand side of Eq.~(\ref{EqImLim}), whereas the second integral disappears for large $t$ due to the Riemann-Lebesgue theorem. This proves limits~(\ref{EqImLim}) and~(\ref{EqZetaLim}). For the Drude--Lorentz spectral density, from Eq.~(\ref{EqgDL}), we see that the convergence is exponential with the rate $\Omega$.

In Fig.~\ref{FigSpinBosonCoh}, we see a very good agreement of formula~(\ref{EqCohSum}) with the numerically exact results.

\subsection{Rate of decoherence and range of validity of the approximation}\label{SecDecoh}

Now let us consider the initial system-bath state of the form
\begin{equation}
\rho(0)=\rho_S(0)\otimes\rho_B,
\end{equation}
where $\rho_S(0)$ is an arbitrary (not necessarily diagonal anymore) initial system state. Substitution of Eq.~(\ref{EqQ1Inhom}) into Eq.~(\ref{EqCoh}) gives
\begin{equation}\label{EqCohSumFull}
\rho_{nm}(t)=\rho_{nm}^{(\rm decoh)}+
\rho_{nm}^{(\rm eq)}(t)+\rho_{nm}^{(\rm noneq)}(t)
+\rho_{nm}^{(\rm coh-coh)}(t),
\end{equation}
where 
\begin{equation}\label{EqDecoh}
\begin{split}
\rho_{nm}^{(\rm decoh)}(t)&=
\rho_{nm}(0)\zeta_{mnn}(t,0)e^{i(\varepsilon_m-\varepsilon_n)t}
\\
&=
\rho_{nm}(0)e^{i(\varepsilon_m-\varepsilon_n)t}
\\
\times
\exp
\Big\{
-&\sum_{\alpha\beta}
(\theta_{\alpha n}-\theta_{\alpha m})
[\theta_{\beta n}g_{\alpha\beta}(t)-\theta_{\beta m}g^*_{\alpha\beta}(t)]
\Big\}
\end{split}
\end{equation}
describes the decoherence in the basis $\{\ket n\}$
(note that $\zeta_{mnn}(t,\tau)$ is independent of $\tau$) and
\begin{multline}
\rho_{nm}^{(\rm coh-coh)}(t)
\\=
i\sum_{l\neq n,m}\int_0^td\tau\,
\Big\{
\rho_{nl}(0)J_{lm}\zeta_{nml}^*(t,\tau)
e^{i[\varepsilon_m\tau-\varepsilon_nt
+\varepsilon_l(t-\tau)]}
\\
-
\rho_{lm}(0)J_{nl}\zeta_{mnl}(t,\tau)
e^{i[\varepsilon_mt-\varepsilon_n\tau
-\varepsilon_l(t-\tau)]}
\Big\}
\end{multline}
describes the coherence--coherence transfer.

Note also that $\mathcal Q\rho(0)\neq0$ gives contributions also to the initial dynamics of $\mathcal P\rho(t)$ (i.e., populations in our case), see Eq.~(\ref{EqMasterInhom}) and Refs.~\cite{Seibt,TrushJCP}. However, here, we neglect this influence.

In this subsection, we focus on the decoherence term. Let us consider the initial state
\begin{equation}\label{EqIniCoh}
\rho(0)=
\left(
\cos\frac\pi8\ket1+\sin\frac\pi8\ket0
\right)
\left(
\cos\frac\pi8\bra1+\sin\frac\pi8\bra0
\right)\otimes\rho_B
\end{equation}
for our example.
From one side, it has initial coherences in the pointer basis $\{\ket+,\ket-\}$. From the other side, the populations in this basis are not stationary. The other parameters are the same. 

\begin{figure}[h]
\begin{centering}
\includegraphics[width=\linewidth]{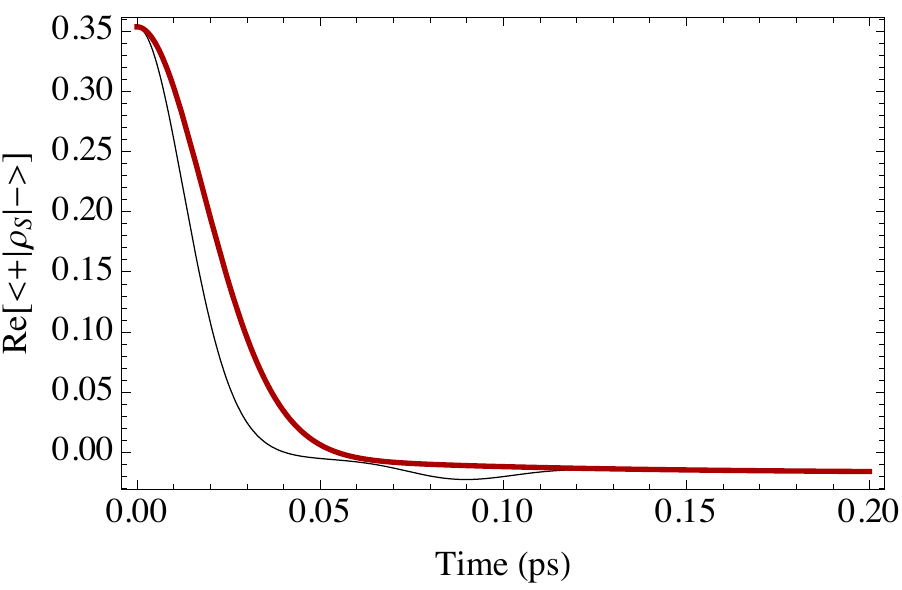}
\vskip -2mm
\caption
{\small Decoherence for the spin-boson model at ultrastrong coupling: formula~(\ref{EqCohSumFull}) (thick red line) in comparison with the numerically exact method of HEOM (thin black line) for the initial state~(\ref{EqIniCoh}).
}
\label{FigDecoh}
\end{centering}
\end{figure}

In Fig.~\ref{FigDecoh}, we again compare our approximation [formula~(\ref{EqCohSumFull})] and the numerically exact solution and show a good precision of formula~(\ref{EqCohSumFull}) for our case. Our approximation slightly underestimates the rate of decoherence. Probably, the cause is that formula~(\ref{EqDecoh}) takes into account the pure decoherence only due to the dynamics of differently displaced baths and does not take into account the decoherence due to the transitions in the basis $\{\ket+,\ket-\}$. According to the used approximation, the transitions on the short time of decoherence are negligible.

\begin{figure}[h]
\begin{centering}
\includegraphics[width=\linewidth]{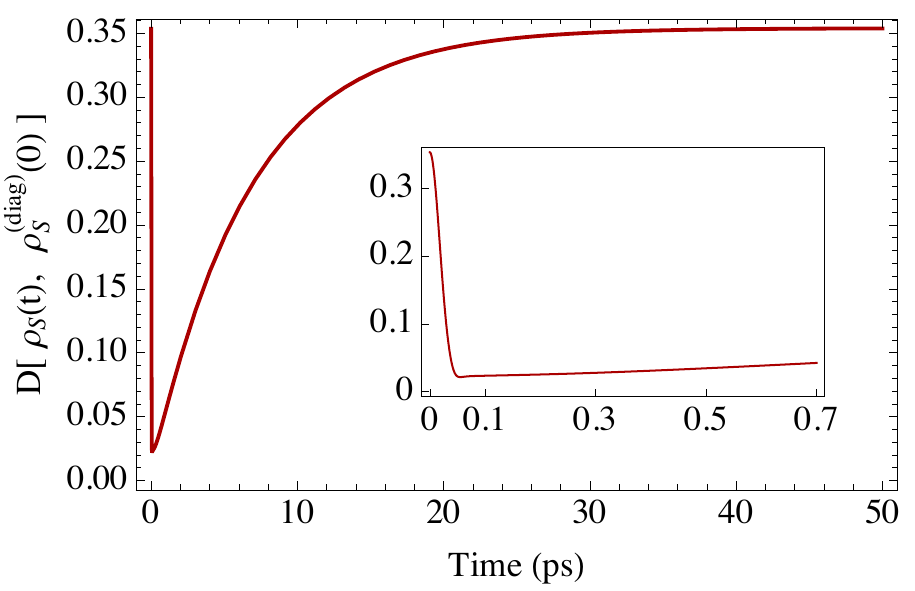}
\vskip -2mm
\caption
{\small For the example in Fig.~\ref{FigDecoh}, the trace distance of $\rho_S(t)$ to the diagonal part $\rho_S^{(\rm diag)}(0)$ of the initial system state $\rho_S(0)$, Eq.~(\ref{EqRhoSd}).
}
\label{FigDist}
\end{centering}
\end{figure}

In Fig.~\ref{FigDist}, we show the trace distance of $\rho_S(t)$ to the diagonal part of $\rho_S(0)$, i.e., to $\rho_S^{(\rm diag)}(0)$, where
\begin{equation}\label{EqRhoSd}
\begin{split}
\rho_S^{(\rm diag)}(t)&=\ket+\braket{+|\rho_S(t)|+}\bra+
\\
&+
\ket-\braket{-|\rho_S(t)|-}\bra-.
\end{split}
\end{equation}
We clearly see two time scales: rapid decoherence to a state close to the projection $\mathcal P\rho(0)$ and further slow evolution toward the steady state. 

This may resolve a discussion in the papers~\cite{Kawai,CresserAnders} about the correct form of the steady state at ultrastrong coupling. Namely, the projection $\mathcal P\rho(0)$ is quasi-steady and becomes exact steady state in the quantum Zeno limit ($\theta_{\alpha n}\to\infty$ for all $\alpha$ and $n$), or, in other words, in the limit of not even ultrastrong but infinitely strong coupling limit. However, in the case of finitely large interaction, it is quasi-steady and the mean force Gibbs state is the only true steady state. The rate of convergence to the mean force Gibbs state decreases to zero when the system-bath interaction strength indefinitely increases.

\begin{figure}[h]
\begin{centering}
\includegraphics[width=\linewidth]{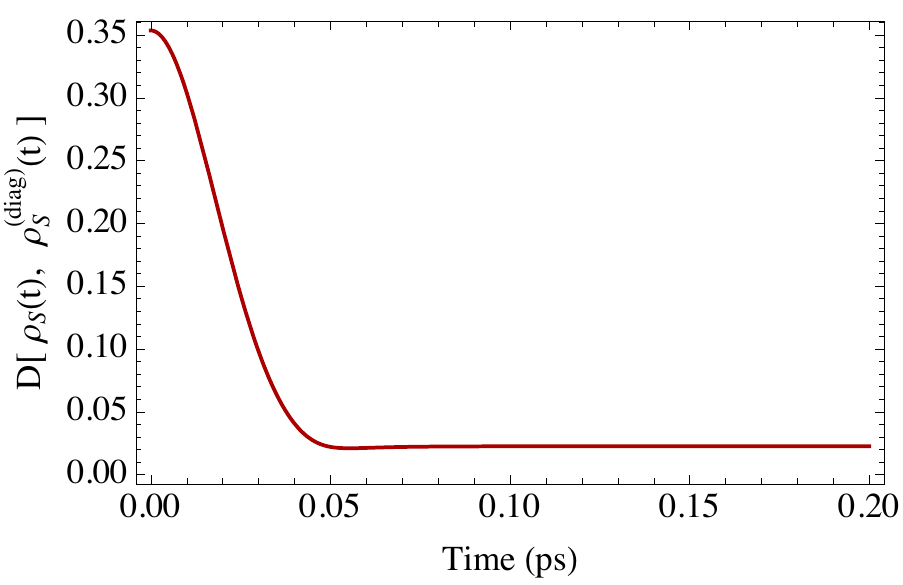}
\vskip -2mm
\caption
{\small For the example in Fig.~\ref{FigDecoh}, the trace distance of $\rho_S(t)$ to the time-dependent diagonal part $\rho_S^{(\rm diag)}(t)$, Eq.~(\ref{EqRhoSd}).
}
\label{FigDistT}
\end{centering}
\end{figure}

For comparison, Fig.~\ref{FigDistT} shows the trace distance of $\rho_S(t)$ to the time-dependent diagonal part $\rho_S^{(\rm diag)}(t)$. It is another illustration of the rapid relaxation toward the subspace $\mathcal{PS}$ (see the beginning of Sec.~\ref{SecProjForst}) and the further slow evolution in the neighborhood of this subspace.

Now we can discuss the range of validity of the presented approach. The approach is essentially based on the described time separation: Evolution in the neighborhood of the subspace $\mathcal{PS}$ should be much slower than the relaxation toward this subspace. As we described in Sec.~\ref{SecProjForst}, relaxation toward this subspace has two parts: the decoherence and the displaced bath relaxation. 

The rate of the displaced bath relaxation can be associated with the rate of convergence of limit~(\ref{EqZetaLim}). In the case of the Drude--Lorentz spectral density (\ref{EqDL}), it is equal to $\Omega$. 

The rates of decoherence are given by the quantities
\begin{equation}\label{EqDecohQuantity}
\sum_{\alpha\beta}(\theta_{\alpha n}-\theta_{\alpha m})
(\theta_{\beta n}-\theta_{\beta m})
\Re[g_{\alpha\beta}(t)],
\end{equation}
from Eq.~(\ref{EqDecoh}). The same quantity enters $\zeta_{nm}$, which, according to  Eq.~(\ref{EqCohEq}), defines the magnitude of coherences. The magnitude of coherences should be small for the validity of the approach. For $t\gg\Omega^{-1}$, expression (\ref{EqDecohQuantity}) is approximately equal to $r_{nm}^{(\rm decoh)}t$, where
\begin{equation}\label{EqDecohRate}
r_{nm}^{(\rm decoh)}=
\sum_{\alpha\beta}
(\theta_{\alpha n}-\theta_{\alpha m})
(\theta_{\beta n}-\theta_{\beta m})
\frac{2\eta}{\beta\Omega}.
\end{equation}

Thus, the transition rates $\gamma_{nm}$ describing the evolution in the neighborhood of $\mathcal{PS}$, should be smaller than both $\Omega$ and Eq.~(\ref{EqDecohRate}). This condition is satisfied either for small $J_{nm}$ or for large system-bath couplings $\eta\theta_{\alpha n}$ (i.e., for the ultrastrong coupling).

Also we see that, for all $n\neq m$, the difference $|\theta_{\alpha n}-\theta_{\alpha m}|$ should not be small at least for some $\alpha$. We assumed that the non-zero $\theta_{\alpha n}$ are not degenerate. So, we should assume also that they are even not quasi-degenerate. The case of degenerate and quasi-degenerate $\theta_{\alpha n}$ is analyzed in  the next section.

\section{Strong-decoherence approximation: General case}
\label{SecGen}

\subsection{Decomposition of the Hamiltonian}

In this section we describe the strong-decoherence approximation in the general case. First, we release assumptions (i) and (ii) in Sec.~\ref{SecForst}. Namely, we allow for the case
\begin{equation}\label{EqAPi}
A_\alpha=\sum_n \theta_{\alpha n}\Pi_n,
\end{equation}
where $\Pi_n$ are orthogonal projectors such that $\sum_n\Pi_n=I_S$ (the identity operator in $\mathcal H_S$), but not necessarily one-dimensional projectors. Without loss of generality, we assume that, for each $\alpha$, all $\theta_{\alpha n}$ are different.

Second, we allow for small correction to Eq.~(\ref{EqAPi}). For example, in the end of the previous section, a possibility of quasi-degeneracies in $\theta_{\alpha n}$ is mentioned. They can be expressed as sums of exactly degenerate $\theta_{\alpha n}$ and small corrections. Moreover, we allow for small corrections of a more general form, not necessarily of the decoherence type. Namely, we decompose of the interaction Hamiltonian in the following way:
\begin{subequations}\label{EqHIfull}
\begin{eqnarray}
H_I&=&\sum_{\alpha=1}^M A_\alpha\otimes B_\alpha
\nonumber
\\
&=&\sum_n
\sum_{\alpha=1}^M
\theta_{\alpha n}\Pi_n\otimes B_\alpha
 +\sum_{\alpha=1}^M \delta A_\alpha\otimes B_\alpha,
 \nonumber
\end{eqnarray}
where
\begin{equation}
\delta A_\alpha=A_\alpha-
\sum_n\theta_{\alpha n}\Pi_n.
\end{equation}
Since
\begin{equation*}
\delta A_\alpha=
\sum_n\Pi_n\delta A_\alpha\Pi_n
+\sum_{n\neq m}\Pi_n A_\alpha\Pi_m,
\end{equation*}
we can write
\begin{eqnarray}
H_I&=&\sum_n
\sum_{\alpha=1}^M
\theta_{\alpha n}\Pi_n\otimes B_\alpha
\label{EqHIdec}
\\
&+&
 \sum_{n\neq m}
 \sum_{\alpha=1}^M \Pi_n A_\alpha\Pi_m\otimes B_\alpha
\label{EqHIod}
\\
&+&
\sum_n
\sum_{\alpha=1}^M
\Pi_n\delta A_\alpha\Pi_n\otimes B^{(n)}_\alpha
\label{EqHId}
\\
&-&
\sum_n
\sum_{\alpha=1}^M
(\delta\varepsilon_{\alpha n}+\delta\varepsilon_{n\alpha})
\Pi_n\delta A_\alpha\Pi_n,
\label{EqHIreorg}
\end{eqnarray}
\end{subequations}
where we have introduced the displaced operators
\begin{eqnarray}
B_\alpha^{(n)}&=&\int [\overline{d_\alpha(\xi)}a_n(\xi)+d_\alpha(\xi)a_n(\xi)^\dag]\,d\xi
\nonumber\\
&=&B_\alpha+\sum_{\beta=1}^M\theta_{\beta n}
(\delta\varepsilon_{\alpha \beta}+\delta\varepsilon_{\beta\alpha})
\label{EqBdisp}
\end{eqnarray}
such that 
\begin{equation}
\Tr\rho_B^{(n)}B^{(n)}_\alpha=0.
\end{equation} 

Term (\ref{EqHIdec}) describes pure decoherence between the subspaces $\mathcal H_S^{(n)}=\Pi_n\mathcal H_S$. It is assumed to be large, thus giving the name for the approximation (the strong-decoherence approximation). Term (\ref{EqHIreorg}) can be assigned to the system Hamiltonian and, thus, is not required to be small. 

Terms (\ref{EqHIod}) and (\ref{EqHId}) are assumed to be small. Term (\ref{EqHIod}) is responsible for transitions between different subspaces $\mathcal H_S^{(n)}$, along with the off-diagonal terms of the system Hamiltonian
\begin{equation}
V=\sum_{n\neq m}\Pi_n H_S\Pi_m.
\end{equation}
Term (\ref{EqHId}) is responsible for the weak-coupling dynamics inside each subspace.

Thus, we have the following decomposition of the Hamiltonian into a reference part $H_0$ and a small perturbation $H'$:
\begin{equation}\label{EqDecomp0}
H=H_0+H',
\end{equation}
where 
\begin{equation*}
\begin{split}
H_0&=\sum_n\Big\{\Pi_n H_S\Pi_n
-\sum_{\alpha=1}^M
(\delta\varepsilon_{\alpha n}+\delta\varepsilon_{n\alpha})
\Pi_n\delta A_\alpha\Pi_n\Big\}
\\
&+H_B+\sum_n\sum_{\alpha=1}^M\theta_{\alpha n}B_\alpha
\end{split}
\end{equation*}
and
\begin{gather}
H'=V+\sum_{n}H_I^{(n)}+
\sum_{n\neq m}\Pi_n A_\alpha\Pi_m\otimes B_\alpha,
\label{EqHprime}
\\[2ex]
H^{(n)}_I=\Pi_n\delta A_\alpha\Pi_n\otimes B^{(n)}_\alpha.
\label{EqHIn}
\end{gather}
We again introduce the Hamiltonian of displaced oscillators $H_B^{(n)}$ (\ref{EqHBn}) so that
\begin{equation}\label{EqHBnDecompose}
H_B^{(n)}=H_B+\sum_{\alpha=1}^M\theta_{\alpha n}B_\alpha+
\delta\varepsilon_n
\end{equation}
and express
\begin{equation}\label{EqH0}
H_0=\sum_{n}
(\bar H_S^{(n)}+H_B^{(n)}),
\end{equation}
where
\begin{eqnarray}
\bar H_S^{(n)}&=&\Pi_n H_S\Pi_n
-
\delta\varepsilon_n\Pi_n
\nonumber
\\&-&
\sum_{\alpha=1}^M
(\delta\varepsilon_{\alpha n}+\delta\varepsilon_{n\alpha})
\Pi_n\delta A_\alpha\Pi_n.
 \label{EqHSn}
\end{eqnarray}

\subsection{Projection operator}
\label{SecProj}

Again, the unperturbed dynamics governed by $H_0$ leads to fast decoherence with respect to the subspaces $\mathcal H_S^{(n)}$, hence, the density operator quickly becomes block-diagonal: $\rho=\sum_n\rho^{(n)}$, where $\rho^{(n)}\in\mathcal S^{(n)}=\Pi_n\mathcal S\Pi_n$. Also, again, from  Eqs.~(\ref{EqH0}) and~(\ref{EqHSn}), we see that the fast dynamics in the subspaces $\mathcal S^{(n)}$ as well as that the dynamics of the system and the bath  inside each subspace are decoupled from each other. Since, inside each subspace, the bath quickly thermalizes, we can define the projection operator as Eq.~(\ref{EqP}), where $\Pi_n$ is now, in general, multidimensional.

\begin{figure}[h]
\begin{centering}
\includegraphics[scale=1.3]{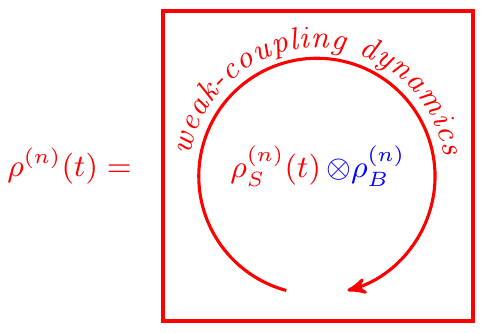}
\vskip -2mm
\caption
{\small
Modification of the bottom part of Fig.~\ref{FigScheme} for the general strong-decoherence coupling. Now $\Pi_n$ are projectors onto, in general, multidimensional system subspaces. The dimensions are not necessarily equal. The blue color denotes the fast processes and the red color denotes the slow processes and degrees of freedom, which are described by a quantum master equation. Here, $\rho^{(n)}(t)=\Pi_n\rho(t)\Pi_n$ and $\rho_S^{(n)}(t)=\Pi_n\rho_S(t)\Pi_n$. The weak-coupling dynamics consists of the fast bath relaxation toward the (displaced) equilibrium state $\rho_B^{(n)}$ and the slow dynamics of $\rho_S^{(n)}(t)$.}
\label{FigSchemeGen}
\end{centering}
\end{figure}

The slow dynamics consists of the dynamics inside each subspace according to the weak coupling theory and transitions between different subspaces. The weak coupling dynamics inside each subspace is defined by the Hamiltonian
\begin{equation}
H^{(n)}=\bar H_S^{(n)}+H_B^{(n)}+H_I^{(n)}.
\end{equation}
and the bath equilibrium state $\rho_B^{(n)}$. The transitions between different subspaces are governed by the off-diagonal blocks of $H'$. Fast and slow dynamics can be again schematically represented by Fig.~\ref{FigScheme} with a slightly modified bottom part for $\rho^{(n)}=\Pi_n\rho(t)\Pi_n$ given in Fig.~\ref{FigSchemeGen}.

\subsection{Particular cases}
\label{SecCases}

\subsubsection{Generalized F\"orster theory}
\label{SecForstGen}

In Sec.~\ref{SecSimp}, we introduced a simple (non-degenerate) version of the strong-decoherence approximation as a generalization of the F\"orster approximation. The presented version with multidimensional projectors $\Pi_n$ includes the generalized F\"orster theory \cite{QEffBio}. In biological light-harvesting complexes one often faces with the case of weakly coupled clusters of molecules. But the dipole couplings between the molecules inside each cluster are not small. The generalized F\"orster theory describes transitions between the cluster. The dynamics inside each cluster can be described in various approximations \cite{Banchi}, including the Redfield (weak coupling) approximation \cite{Renger2008}, which is our case. 

Note that, in Ref.~\cite{Jang2014}, another projection operator is proposed for the generalized F\"orster theory:
\begin{equation}
\mathcal P\rho=\sum_{n}p_n Z_n^{-1}e^{-\beta H^{(n)}},
\end{equation}
where $Z_n=\Tr e^{-\beta H^{(n)}}$ and $p_n=\Tr\Pi_n\rho$.
This means that the system-bath coupling inside each cluster is not weak and the system and the bath thermalize together to their ``global'' thermal state, i.e., with respect to $H^{(n)}$. This thermalization process is treated as ``fast''. In our approximation, the the system-bath coupling inside each cluster is weak, so, the bath alone thermalize much faster than the system, which leads to projection operator (\ref{EqP}).

\subsubsection{Modified Redfield theory}
\label{SecmRedf}

Consider again the EET Hamiltonian $H_S$ (\ref{EqHSforst}), where $\ket n$ corresponds to an excitation on molecule $n$, and $A_\alpha\equiv A_n=\ket n\bra n$. Let the spectrum $H_S$ is non-degenerate and the intersite couplings $J_{nm}$ are small compared to the local excitation energies $\varepsilon_n$. Then the eigenvectors $\ket{e_n}$ of $H_S$  are highly localized: $\braket{e_n|m}$ are small if $n\neq m$. Put $\Pi_n=\ket{e_n}\bra{e_n}$. Then $V=0$ and [see Eq.~(\ref{EqHprime})]
\begin{equation}
H'=\sum_k
\sum_{n\neq m}
\braket{e_n|k}\braket{k|e_m}
\ket{e_n}\bra{e_m}\otimes B_\alpha
\end{equation}
is the
off-diagonal of the interaction Hamiltonian in the eigenbasis of the system Hamiltonian.
For each $k$, at least one of scalar products $\braket{e_n|k}$ and $\braket{k|e_m}$ is small. Hence, $H'$  can be treated as a small perturbation even if the system-bath coupling is large. In theory of EET, this approximation is referred to as the modified Redfield approximation, in contrast to the usual (``standard'') Redfield approximation (or the weak-coupling approximation), where the whole interaction Hamiltonian is treated perturbatively.

So, the range of validity of the modified Redfield approach intersects with that of the F\"orster approach. The difference is that the F\"orster approach considers the decoherence in the local basis as the primary process, while the modified Redfield --- the decoherence in the eigenbasis.

From another side, the usual weak coupling approach also falls into the modified Redfield approach whenever all energy levels are non-degenerate and well separated from each other. This ensures the fast decoherence despite of the fact that the diagonal part of $H_I$ is also weak. 

The modified Redfield theory fails if there are some degenerate or nearly degenerate energy levels \cite{NovoGrond,NovoGrond2013,NovoGrond2017}. This was considered as a technical limitation, which can be overcome by the inclusion of non-secular terms (pumping of coherences from the populations). However, in Ref.~\cite{TrushJCP}, where such terms were introduced, it is argued that this limitation is fundamental since the basic assumption of strong decoherence between different eigenvectors is not satisfied.

Hence, a hybrid consideration where the dynamics inside the subspaces of degenerate or nearly degenerate levels is described in the weak coupling approximation, while the transitions between these subspaces is treated using the modified Redfield approach, can be useful. This falls into the proposed general formalism. 

Combinations of the modified Redfield and the F\"orster approaches are also widely used in theory of EET \cite{NovoGrond,Yang2003,Renger2011,NovoGrond2017} and also can be considered within the proposed formalism.

\subsection{Master equation}

Substitution of interaction Hamiltonian (\ref{EqHprime}) and projection operator (\ref{EqP}) to the general Markovian master equation (\ref{EqMaster}) gives

\begin{equation}\label{EqMasterGen}
\dot\rho^{(\rm d)}_S(t)=\sum_n\mathcal R_n\rho^{(\rm d)}_S(t)
+\sum_{m\neq n}
\mathcal T_{nm}\rho^{(\rm d)}_S(t),
\end{equation}
where 
\begin{equation}\label{EqRhoSdiag}
\rho^{(\rm d)}_S=\Tr_B\mathcal P\rho=\sum_n\Pi_n\rho_S\Pi_n\equiv \sum_n\rho_S^{(n)},
\end{equation}
\begin{equation}
\mathcal R_n(t)\rho^{(\rm d)}_S=
\int_0^\infty d\tau\,
\Tr_B\big\{
\mathcal L'_n(t)
\mathcal L'_n(t-\tau)[\rho_S^{(n)}\otimes\rho_B^{(n)}]
\big\}
\end{equation}
is the Redfield generator in the subspace $\mathcal H_S^{(n)}$ with $\mathcal L'_n(t)=[H_I^{(n)},\,\cdot\,]$, while $\mathcal T_{nm}$ come from the off-diagonal contributions to $H'$ and describe the transitions between different subspaces. Namely, for fixed $n$ and $m$, $\mathcal T_{nm}$ describes transitions from the subspace $\mathcal H_S^{(m)}$ to the subspace $\mathcal H_S^{(n)}$.

For completeness, let us give an explicit expression for the Redfield generator:
\begin{multline}
\mathcal R_n\rho^{(\rm d)}_S=-i[H_{\rm LS}^{(n)},\rho^{(\rm d)}_S]
+\sum_{\alpha,\beta=1}^M\sum_{\omega,\omega'\in\mathcal F_n}
e^{i(\omega'-\omega)t}
\\
\times
\gamma_{\alpha\beta}(\omega,\omega')\Big(
A_{\beta n\omega}\rho^{(\rm d)}_SA_{\alpha n\omega'}^\dag
-\frac12\big\{
A_{\alpha n\omega'}^\dag A_{\beta n\omega},
\rho^{(\rm d)}_S
\big\}
\Big),
\end{multline}
where
\begin{equation}\label{EqHLSRedf}
H_{\rm LS}^{(n)}=
\sum_{\alpha,\beta=1}^M\,
\sum_{\omega,\omega'\in\mathcal F_n}
e^{i(\omega'-\omega)t}
S_{\alpha\beta}(\omega,\omega')
A_{\alpha n\omega'}^\dag A_{\beta n\omega}
\end{equation}
(the subindex LS stands for the Lamb shift). Here $\mathcal F_n$ is the spectrum of $[\bar H_S^{(n)},\,\cdot\,]$, or, in other words, the set of all Bohr frequencies (all differences between eigenvalues) of $\bar H_S^{(n)}$. Note that $\mathcal F_n$ includes positive and negative Bohr frequencies and the zero Bohr frequency. Denote $\spec \bar H_S^{(n)}$ the spectrum of  $\bar H_S^{(n)}$ and $P_\varepsilon$ the projector onto the eigenspace corresponding to $\varepsilon\in\spec\bar H_S^{(n)}$. Put $P_\varepsilon\equiv0$ whenever $\varepsilon\notin\spec\bar H_S^{(n)}$. Then, $A_{\alpha n}=\Pi_n\delta A_\alpha\Pi_n$, 
\begin{equation}
A_{\alpha n\omega}=\sum_{\varepsilon\in\spec\bar H_S^{(n)}}
P_{\varepsilon-\omega}A_{\alpha n}P_\varepsilon,
\end{equation}
so that $[\bar H_S^{(n)},A_{\alpha n\omega}]=-\omega A_{\alpha n\omega}$. Also,
\begin{gather}
\begin{aligned}
\gamma_{\alpha\beta}(\omega,\omega')&=
\Gamma_{\alpha\beta}(\omega)+\Gamma^*_{\beta\alpha}(\omega'),\\
S_{\alpha\beta}(\omega,\omega')&=
\frac1{2i}\left[\Gamma_{\alpha\beta}(\omega)-\Gamma^*_{\beta\alpha}(\omega')
\right]
\end{aligned}\nonumber
\\
\Gamma_{\alpha\beta}(\omega)=
\int_0^\infty d\tau\,e^{i\omega \tau}
\langle e^{iH_B\tau}B_\alpha e^{-iH_B\tau}B_\beta\rangle.
\end{gather}

The intersubspace transition superoperators $\mathcal T_{nm}$ are derived completely analogously:

\begin{multline}
\mathcal T_{nm}\rho^{(\rm d)}_S
=-i[H_{\rm LS}^{(nm)},\rho^{(\rm d)}_S]
\\+
\sum_{\alpha,\beta=0}^M\,
\sum_{\omega,\omega'\in\mathcal F_{nm}}
e^{i(\omega'-\omega)t}
\gamma_{\alpha\beta nm}(\omega,\omega')
\\
\times\Big(
A_{\beta nm\omega}\rho^{(\rm d)}_S
A_{\alpha nm\omega'}^\dag
-\frac12
\big\{
A_{\alpha nm\omega'}^\dag
A_{\beta nm\omega},\rho^{(\rm d)}_S
\big\}
\Big),
\end{multline}
where
\begin{multline}\label{EqHLST}
H_{\rm LS}^{(nm)}=
\sum_{\alpha,\beta=0}^M\,
\sum_{\omega,\omega'\in\mathcal F_{nm}}
e^{i(\omega'-\omega)t}
\\
\times
S_{\alpha\beta nm}(\omega,\omega')
A_{\alpha nm\omega'}^\dag A_{\beta nm\omega}.
\end{multline}
Here, we have formally put $A_0\equiv V$ and $B_0\equiv I_B$ (the identity operator in the bath Hilbert space).
Further, $\mathcal F_{nm}$ is the set of differences $\varepsilon'-\varepsilon$, where $\varepsilon'\in\spec\bar H_S^{(m)}$ and $\varepsilon\in\spec\bar H_S^{(n)}$. So, the union of all $\mathcal F_n$ and all $\mathcal F_{nm}$ is the set of the Bohr frequencies of the diagonal part of the system Hamiltonian
\begin{equation}
\bar H_S^{\text{(d)}}=\sum_n \bar H_S^{(n)}.
\end{equation}
Then, $A_{\alpha nm}=\Pi_n A_\alpha\Pi_m$ and
\begin{equation}
A_{\alpha nm\omega}=\sum_{\varepsilon\in\spec\bar H_S^{(m)}}
P_{\varepsilon-\omega}A_{\alpha nm}P_\varepsilon,
\end{equation}
so that $[\bar H_S^{\text{(d)}},A_{\alpha nm\omega}]=-\omega A_{\alpha nm\omega}$. Also,
\begin{equation}\label{EqgS}
\begin{split}
\gamma_{\alpha\beta nm}(\omega,\omega')&=
\Gamma_{\alpha\beta nm}(\omega)+\Gamma^*_{\beta\alpha nm}(\omega'),\\
S_{\alpha\beta nm}(\omega,\omega')&=
\frac1{2i}\left[\Gamma_{\alpha\beta nm}(\omega)-\Gamma^*_{\beta\alpha nm}(\omega')
\right],
\end{split}
\end{equation}
\begin{eqnarray}
\Gamma_{\alpha\beta nm}(\omega)&=&
\int_0^\infty d\tau\,e^{i\omega \tau}
\langle e^{iH_B^{(m)}\tau} B_\alpha e^{-iH_B^{(n)}\tau} B_\beta\rangle_m
\nonumber
\\
&\equiv&
\int_0^\infty d\tau\,e^{i\omega \tau}
\zeta_{\alpha\beta nm}(\tau).
\label{EqG}
\end{eqnarray}
Note that the previously defined functions $\zeta_{nm}(\tau)$ [see Eq.~(\ref{EqIntegrand})] coincide with $\zeta_{00nm}(\tau)$.

Though the generator looks like the Gorini--Kossakowski--Lindblad--Sudarshan (GKLS) form, it is not of the GKLS form, because the matrices $\gamma_{\alpha\beta}(\omega,\omega')$ and $\gamma_{\alpha\beta nm}(\omega,\omega')$ [with two double indexes $i=(\alpha,\omega')$ and $j=(\beta,\omega)$] are, in general, not positive-semidefinite. However, if all exponents $e^{i(\omega'-\omega)t}$ for $\omega'\neq\omega$ can be treated as rapidly oscillating, then we can drop all the terms with $\omega'\neq\omega$ (secular approximation). Then, for each $\omega$, the matrices $\gamma_{\alpha\beta}(\omega,\omega)$ and $\gamma_{\alpha\beta nm}(\omega,\omega)$ (with the simple indices $\alpha$ and $\beta$) are positive-semidefinite. Hence, the master equation becomes of the first standard GKLS form \cite{BP}.

If the secular approximation cannot be applied, then another approximation [namely, a partial secular approximation and small modifications in the arguments of rate constants $\gamma_{\alpha\beta}(\omega)$ and $\gamma_{\alpha\beta nm}(\omega)$] can be applied to obtain a master equation in the GKLS form \cite{Uni}.


\subsection{Mean-force Gibbs steady state}

In the considered limit, the mean force Gibbs state tends to
\begin{equation}\label{EqSt}
\rho^{(\rm st)}_S=Z_S^{-1}e^{-\beta \bar H_S^{(\rm d)}},
\quad Z_S=\Tr e^{-\beta \bar H_S^{(\rm d)}}.
\end{equation}
Let us prove that this state is stationary for the master equation derived in the previous subsection if we apply the secular approximation (also described in the end of the previous subsection). Denote the corresponding generators $\mathcal R_n^{(\rm sec)}$ and $\mathcal T_{nm}^{(\rm sec)}$.

The projection of state (\ref{EqSt})  onto the subspace $\mathcal H_S^{(n)}$ gives
\begin{equation}
\Pi_n\rho^{(\rm st)}_S\Pi_n=Z_S^{-1}e^{-\beta \bar H_S^{(n)}},
\end{equation}
i.e., the Gibbs state with respect to $\bar H_S^{(n)}$ (up to a normalization constant). It is well-known to be stationary for the secular Redfield generator $\mathcal R_n^{(\rm sec)}$. So, it suffices to prove the stationarity of state (\ref{EqSt}) for the transition part  $\sum\mathcal T_{nm}^{(\rm sec)}$ of the generator.

Since $[A_{\alpha nm\omega}^\dag A_{\alpha nm\omega},\bar H_S^{(\rm d)}]=0$, state (\ref{EqSt}) commutes with the Lamb-shift Hamiltonian (\ref{EqHLST}) in the secular approximation. Now establish the detailed balance conditions
\begin{equation}\label{EqDetBal}
\gamma_{\alpha\beta nm}(\omega)=\gamma_{\beta\alpha mn}(-\omega)e^{\beta\omega},
\end{equation}
where 
\begin{multline}\label{Eqgammasec}
\gamma_{\alpha\beta nm}(\omega)\equiv \gamma_{\alpha\beta nm}(\omega,\omega)\\=
\int_{-\infty}^\infty d\tau\,e^{i\omega \tau}
\langle e^{iH_B^{(m)}\tau} B_\alpha e^{-iH_B^{(n)}\tau} B_\beta\rangle_m.
\end{multline} 
As in Sec.~\ref{SecStForst}, let us change the variable of integration in Eq.~(\ref{Eqgammasec}) by $s=-\tau-i\beta$:
\begin{multline*}
\gamma_{\alpha\beta nm}(\omega)=
e^{\beta\omega}
\int_{-\infty}^\infty d\tau\,e^{-i\omega s}
\\\times
\langle e^{-iH_B^{(m)}s} 
B_\alpha
e^{iH_B^{(n)}s} 
e^{-\beta H_B^{(n)}} 
B_\beta
e^{\beta H_B^{(m)}}
\rangle_m.
\end{multline*} 
Now applying Eq.~(\ref{EqOmn}), we obtain
\begin{multline*}
\gamma_{\alpha\beta nm}(\omega)=
e^{\beta\omega}
\int_{-\infty}^\infty d\tau\,e^{-i\omega s}
\\\times
\big\langle e^{-iH_B^{(m)}s} 
B_\alpha
e^{iH_B^{(n)}s} 
\big(
e^{-\beta H_B^{(n)}} 
B_\beta
e^{\beta H_B^{(n)}}
\big)
\big\rangle_n.
\end{multline*} 
Now we apply the Kubo--Martin--Schwinger condition
\begin{equation}
\big\langle
Y
\big(
e^{-\beta H_B^{(n)}} 
X
e^{\beta H_B^{(n)}}
\big)
\big\rangle_n
=
\langle
XY
\rangle_n
\end{equation}
for $Y=e^{-iH_B^{(m)}s} B_\alpha e^{iH_B^{(n)}s}$ and $X=B_\beta$, which gives
\begin{equation*}
\begin{split}
\gamma_{\alpha\beta nm}(\omega)&=
e^{\beta\omega}
\int_{-\infty}^\infty d\tau\,e^{-i\omega s}
\langle 
B_\beta
e^{-iH_B^{(m)}s} 
B_\alpha
e^{iH_B^{(n)}s} 
\rangle_n
\\
&=
e^{\beta\omega}
\int_{-\infty}^\infty d\tau\,e^{-i\omega s}
\langle 
e^{iH_B^{(n)}s} 
B_\beta
e^{-iH_B^{(m)}s} 
B_\alpha
\rangle_n
\\&=\gamma_{\beta\alpha mn}(-\omega)e^{\beta\omega}.
\end{split}
\end{equation*} 

Now, in view of this detailed balance condition and 
\begin{equation}
A_{\beta nm\omega}\rho^{(\rm st)}_S=
\rho^{(\rm st)}_SA_{\beta nm\omega}e^{-\beta\omega},
\end{equation}
the terms 
$$\gamma_{\alpha\beta nm}(\omega)A_{\beta nm\omega}\rho^{(\rm st)}_SA_{\alpha nm\omega}^\dag$$ 
from $\mathcal T_{nm}^{(\rm sec)}$ are canceled out with the terms $$\frac12\gamma_{\beta\alpha mn}(-\omega)\{A_{\beta mn,-\omega}^\dag A_{\alpha mn,-\omega},\rho^{(\rm st)}_S\}$$ 
from $\mathcal T_{mn}^{(\rm sec)}$ (note that $A_{\beta mn,-\omega}^\dag=A_{\beta nm\omega}$), which proves the stationarity of the mean-force Gibbs state $\rho^{(\rm st)}_S$.

\subsection{Off-diagonal blocks}

We have derived equations for the diagonal blocks $\rho_S^{(\rm d)}$. The off-diagonal blocks (coherences) can be calculated by a slight generalization of methods of Sec.~\ref{SecCoh}.
Denote
\begin{equation}
\begin{split}
\rho_{nm}(t)&=
\Pi_n\Tr_B\{e^{-iH_0t}\rho(t)e^{iH_0t}\}\Pi_m
\\
&=\Pi_n\Tr_B\{e^{-iH_0t}\mathcal Q\rho(t)e^{iH_0t}\}\Pi_m,
\end{split}
\end{equation} 
where again $\mathcal Q=1-\mathcal P$.
If $\rho(0)=\mathcal P\rho(0)$, the the substitution of Eq.~(\ref{EqQ1Pre}) gives
\begin{equation*}
\rho_{nm}(t)\!=\!
-i\!\int_0^t\!\!d\tau\,
\Pi_n\!\Tr_B\{e^{-iH_0t}\mathcal L'(t-\tau)
\mathcal P\rho(t-\tau)e^{iH_0t}\}\Pi_m.
\end{equation*}
Again, we can substitute here $\rho(t-\tau)$ by $\rho(t)$ and obtain
\begin{multline}\label{EqCohGen}
\rho_{nm}(t)=\!
-i\!\int_0^t\!\!d\tau\,
\Pi_n\!\Tr_B\{e^{-iH_0t}\mathcal L'(t-\tau)
\mathcal P\rho(t)e^{iH_0t}\}\Pi_m
\\
=
i\sum_{\alpha=0}^M\sum_{\omega\in\mathcal F_{nm}}
\int_0^td\tau\,e^{i(\bar\varepsilon_m-\bar\varepsilon_n-\omega)\tau}
\\
\times
\big[
\zeta_{0\alpha mn}^*(\tau)
\rho^{(n)}_S(t)A_{\alpha nm\omega}
-
\zeta_{0\alpha nm}(\tau)
A_{\alpha nm\omega}\rho^{(m)}_S(t)
\big].
\end{multline}
The limit $t\to\infty$ and the substitution of $\rho^{(n)}_S(t)$ by the stationary operators $Z_S^{-1}e^{-\beta\bar H_S^{(n)}}$ give the steady-state off-diagonal parts and, thus, correction to the steady-state obtained in the previous subsection.

The non-equilibrium corrections to coherences and the influence of initial coherences also can be calculated analogously to Sec.~\ref{SecCoh}.

\subsection{Degenerate ultrastrong coupling}

For the explicit evaluation of both the rate constants $\Gamma_{\alpha\beta nm}(\omega)$ [Eq.~(\ref{EqG})] and the coherences [Eq.~(\ref{EqCohGen})], we need explicit expressions for the functions $\zeta_{\alpha\beta nm}(\tau)$. For the case of the F\"orster and modified Redfield theories, they are derived in Refs.~\cite{mRedf,YangFl,Seibt} using the cumulant expansion method (which we also use in Appendix~\ref{AppCalc}). The calculation for the considered general case is completely analogous. It is straightforward, but cumbersome.  

For simplicity, we restrict our consideration to the case when $\Pi_n$ are eigenprojectors of $A_\alpha$, i.e., there is no off-diagonal part (\ref{EqHIod}) of $H_I$. In other words, 
\begin{equation}
H'=V+\sum_n H_I^{(n)}.
\end{equation}
This case can be qualified as the degenerate ultrastrong coupling combined with the weak coupling inside the subspaces. This case includes (but not limited to) a hybrid F\"orster--Redfield theory, see Sec.~\ref{SecForstGen}. In this case, we need only the functions $\zeta_{00nm}(\tau)\equiv \zeta_{nm}(\tau)$, which have been already evaluated in Eq.~(\ref{EqIntegrand}).  The expressions generators $\mathcal T_{nm}$ of the master equation (\ref{EqMasterGen}) can be simplified to:

\begin{multline}
\mathcal T_{nm}\rho_S
=-i[H_{\rm LS}^{(nm)},\rho_S]+
\sum_{\omega,\omega'\in\mathcal F_{nm}}
e^{i(\omega'-\omega)t}
\\
\times
\gamma_{nm}(\omega)
\Big(
V_{nm\omega}\rho_S
V_{nm\omega'}^\dag
-\frac12
\big\{
V_{nm\omega'}^\dag
V_{nm\omega},\rho_S
\big\}
\Big),
\end{multline}
where
\begin{equation}\label{EqHLSTV}
H_{\rm LS}^{(nm)}=
\sum_{\omega,\omega'\in\mathcal F_{nm}}
e^{i(\omega'-\omega)t}
S_{nm}(\omega,\omega')
V_{nm\omega'}^\dag V_{nm\omega},
\end{equation}
\begin{gather}
\begin{aligned}
\gamma_{nm}(\omega,\omega')&=
\Gamma_{nm}(\omega)+\Gamma^*_{nm}(\omega'),\\
S_{nm}(\omega,\omega')&=
\frac1{2i}\left[\Gamma_{nm}(\omega)-\Gamma^*_{nm}(\omega')
\right]
\end{aligned}
\label{EqgSV}
\\
\Gamma_{nm}(\omega)=
\int_0^\infty 
\zeta_{nm}(\tau)\,
e^{i(\bar\varepsilon_m-\bar\varepsilon_n-\omega)\tau}
d\tau.
\label{EqGV}
\end{gather}

If we further adopt the secular approximation, then only the terms with $\omega'=\omega$ are present:
\begin{eqnarray}
\gamma_{nm}(\omega,\omega)&\equiv&
\gamma_{nm}(\omega)=2\Re \Gamma_{nm}(\omega)
\nonumber
\\
&=&
\int_{-\infty}^{+\infty}\zeta_{nm}(\tau)\,
e^{i(\bar\varepsilon_m-\bar\varepsilon_n-\omega)\tau}
d\tau,\qquad
\\
S_{nm}(\omega,\omega)&\equiv&
S_{nm}(\omega)=\Im \Gamma_{nm}(\omega).
\end{eqnarray}

Finally,
\begin{multline}\label{EqCohGenV}
\rho_{nm}(t)
=
i\sum_{\omega\in\mathcal F_{nm}}
\int_0^td\tau\,e^{i(\bar\varepsilon_m-\bar\varepsilon_n-\omega)\tau}
\\
\times\big[
\zeta^*_{mn}(\tau)\rho^{(n)}_S(t)V_{nm\omega}
-\zeta_{nm}(\tau)
V_{nm\omega}\rho^{(m)}_S(t)
\big].
\end{multline}

\section{Conclusions}

We have introduced a new regime of evolution of  open quantum systems called the strong-decoherence regime. It includes the ultrastrong-coupling regime as an important particular case. We have derived the corresponding quantum master equations and their steady states, which are equal mean force Gibbs state in the corresponding limit. Also, we have obtained the first-order corrections to these expressions for the steady states.

This formalism can be used for testing theories of strong-coupling quantum thermodynamics \cite{Rivas,DannMegierKosloff}. Thermodynamic of pure decoherence was proposed recently \cite{TDdecoh}. The strong-decoherence approximation can be regarded as a correction to pure decoherence: the strong pure decoherence complemented by the slow transfer between the subspaces.

\begin{acknowledgments}
I am grateful to Janet Anders, James Cresser, Christopher Jarzynski, Camille Lombard Latune, and Alexander Teretenkov for fruitful discussions and useful comments. This work was supported by the Russian Science Foundation (Project No. 17-71-20154).
\end{acknowledgments}

\appendix

\section{Some formulas of the projection operator formalism}
\label{AppPQ}

Denote $\mathcal L=[H,\,\cdot\,]$, where $H=H_0+H'$ is a Hamiltonian, $\mathcal L_0=[H_0,\,\cdot\,]$, $\mathcal L'=[H',\,\cdot\,]$, and $\mathcal L'(t)=[H'(t),\,\cdot\,]$, where $H'(t)=e^{iH_0t}H'e^{-iH_0t}$. Let also a projection (super)operator $\mathcal P$ satisfy $[\mathcal P,\mathcal L_0]=0$ and $\mathcal P\mathcal L'\mathcal P=0$. Denote also $\mathcal Q=1-\mathcal P$. The von Neumann equation for the density operator $\rho(t)$ in the interaction representation with respect to $H_0$ is
\begin{equation}
\dot\rho(t)=-i\mathcal L'(t)\rho(t).
\end{equation}
This is equivalent to the following system of equations:
\begin{subequations}\label{EqPQrho}
\begin{eqnarray}
\mathcal P\dot\rho(t)&=&-i\mathcal P\mathcal L'(t)\mathcal Q\rho(t),\label{EqPrho}\\
\mathcal Q\dot\rho(t)&=&
-i\mathcal L'(t)\mathcal P\rho(t)
-i\mathcal Q\mathcal L'(t)\mathcal Q\rho(t),\label{EqQrho}
\end{eqnarray}
\end{subequations}
If we treat $\mathcal P\rho(t)$ as a known function, then a formal solution of  equation (\ref{EqQrho}) for $\mathcal Q\rho(t)$ is:
\begin{multline}\label{EqQ}
\mathcal Q\rho(t)=
{\rm T}_+\exp\left\lbrace
-i\int_0^t\mathcal Q\mathcal L'(\tau)d\tau
\right\rbrace \mathcal Q\rho(0)\\-
i\int_0^t \,d\tau\,{\rm T}_+\exp
\left\lbrace
-i\int_\tau^t\mathcal Q\mathcal L'(\tau')d\tau'
\right\rbrace\mathcal L'(\tau) \mathcal P\rho(\tau),
\end{multline}
where 
\begin{multline}\label{EqChronExp}
{\rm T}_+\exp\left\lbrace
-i\int_0^t f(\tau)d\tau\right\rbrace\\=
1+\sum_{n=1}^\infty(-i)^n
\int_0^td\tau_1\int_0^{\tau_1}d\tau_2\ldots
\int_0^{\tau_{n-1}}d\tau_n\\
 f(\tau_1)f(\tau_2)\cdots f(\tau_n)
\end{multline}
is the chronological exponential. In particular, if $\mathcal Q\rho(0)=0$, then, in the first order with respect to $\mathcal L'(t)$, we have
\begin{eqnarray}
\mathcal Q\rho(t)&=&
-i\int_0^t \mathcal L'(\tau) \mathcal P\rho(\tau)\,d\tau
\nonumber
\\
&=&
-i\int_0^t \mathcal L'(t-\tau) \mathcal P\rho(t-\tau)\,d\tau.
\label{EqQ1Pre}
\end{eqnarray}
Its substitution into Eq.~(\ref{EqPrho}) gives
\begin{equation}\label{EqMasterFint}
\mathcal P\dot\rho(t)=
-
\int_0^t d\tau\,
\mathcal P\mathcal L'(t)\mathcal L'(t-\tau)\mathcal P\rho(t-\tau).
\end{equation} 
The Markovian approximation is the replacement $\mathcal P\rho(t-\tau)$ by $\mathcal P\rho(t)$ and the extension of the upper limit of integration in Eq.~(\ref{EqMasterFint}) to infinity. Both replacements means that the integrand quickly decays with $\tau$ [much faster than the rate of the evolution of $\mathcal P\rho(t-\tau)$]. Thus, we have Eq.~(\ref{EqMaster}):
\begin{equation}\label{EqMasterInft}
\mathcal P\dot\rho(t)=
-
\int_0^\infty d\tau\,
\mathcal P\mathcal L'(t)\mathcal L'(t-\tau)\mathcal P\rho(t).
\end{equation} 

If $\mathcal Q\rho(0)\neq0$, then formula (\ref{EqQ1Pre}) should be modified:
\begin{eqnarray}
\mathcal Q\rho(t)&=&
\mathcal Q\rho(0)
-i\int_0^t \mathcal Q\mathcal L'(t-\tau) \mathcal Q\rho(0)\,d\tau
\nonumber
\\
&-&i\int_0^t \mathcal L'(t-\tau) \mathcal P\rho(t-\tau)\,d\tau,
\label{EqQ1Inhom}
\end{eqnarray}
if we  take the first-order approximation to the first chronological exponential in Eq.~(\ref{EqQ}). Substitution of Eq.~(\ref{EqQ1Inhom}) to Eq.~(\ref{EqPrho}) gives the master equation for $\mathcal P\rho(t)$ with inhomogeneous terms:
\begin{eqnarray}
\mathcal P\dot\rho(t)=
&-&
\int_0^\infty d\tau\,
\mathcal P\mathcal L'(t)\mathcal L'(t-\tau)
[\mathcal P\rho(t)+\mathcal Q\rho(0)]
\nonumber
\\
&-&i\mathcal P\mathcal L'(t)\mathcal Q\rho(0)
\label{EqMasterInhom}
\end{eqnarray}

\section{Calculation of the rate constants}\label{AppCalc}

Let us derive formula (\ref{EqIntegrand}). We have
\begin{eqnarray}
H_B^{(n)}&=&
H_B+\delta\varepsilon_n+\sum_\alpha\theta_{\alpha n}B_\alpha
\nonumber\\
&=&
H_B^{(m)}+\delta\varepsilon_{n}-\delta\varepsilon_{m}+
\sum_\alpha(\theta_{\alpha n}-\theta_{\alpha m})B_\alpha
\nonumber\\
&=&
H_B^{(m)}+\delta\varepsilon_{n}-\delta\varepsilon_{m}
\nonumber\\
&+&
\sum_\alpha(\theta_{\alpha n}-\theta_{\alpha m})
\Big[B_\alpha^{(m)}-\sum_\beta\theta_{\beta m}(\delta\varepsilon_{\alpha \beta}+\delta\varepsilon_{\beta\alpha})\Big]
\nonumber\\
&=&
H_B^{(m)}+
\sum_{\alpha\beta}
(\theta_{\alpha n}-\theta_{\alpha m})
(\theta_{\beta n}-\theta_{\beta m})
\delta\varepsilon_{\alpha\beta}
\nonumber\\
&+&
\sum_\alpha(\theta_{\alpha n}-\theta_{\alpha m})
B_\alpha^{(m)},
\end{eqnarray}
where the displaced operators $B_\alpha^{(m)}$ were defined in Eq.~(\ref{EqBdisp}).
Thus, 
\begin{multline}\label{EqAppB10}
\langle
e^{iH_B^{(m)}t}e^{-iH_B^{(n)}t}
\rangle_m
\\
=
\Big\langle
e^{iH_B^{(m)}t}
\exp\Big\{
-it\Big[H_B^{(m)}
+\sum_\alpha(\theta_{\alpha n}-\theta_{\alpha m})
B_\alpha^{(m)}\Big]\Big\}
\Big\rangle_m
\\\times
\exp\Big\{
-it\sum_{\alpha\beta}
(\theta_{\alpha n}-\theta_{\alpha m})
(\theta_{\beta n}-\theta_{\beta m})
\delta\varepsilon_{\alpha\beta}
\Big\}
\end{multline}
The first factor can be calculated using the second-order cumulant (Magnus) expansion with respect to $B_\alpha^{(m)}$, which is exact for the bosonic bath due to Wick's theorem \cite{Mukamel}. Since
\begin{multline}
\exp\Big\{
-it\Big[H_B^{(m)}
+\sum_\alpha(\theta_{\alpha n}-\theta_{\alpha m})
B_\alpha^{(m)}\Big]\Big\}
=e^{-iH_B^{(m)}t}
\\\times
{\rm T}_+
\exp\left\lbrace
-i\sum_\alpha(\theta_{\alpha n}-\theta_{\alpha m})
\int_0^t B^{(m)}_\alpha(\tau)\,d\tau
\right\rbrace,
\end{multline}
where 
\begin{equation}
B^{(m)}_\alpha(\tau)=e^{iH_B^{(m)}\tau}B^{(m)}_\alpha e^{-iH_B^{(m)}\tau}
\end{equation} and ${\rm T}_+$ is the chronological exponential (\ref{EqChronExp}), and in view of $\langle B^{(m)}_\alpha(\tau)\rangle_m=0$, the second-order cumulant expansion for the first factor in Eq.~(\ref{EqAppB10}) gives 
\begin{eqnarray}
\Big\langle
&&e^{iH_B^{(m)}t}
\exp\Big\{
-it\Big[H_B^{(m)}
+\sum_\alpha(\theta_{\alpha n}-\theta_{\alpha m})
B_\alpha^{(m)}\Big]\Big\}
\Big\rangle_m
\nonumber
\\
&&=
\bigg\langle
{\rm T}_+
\exp\left\lbrace
-i\sum_\alpha(\theta_{\alpha n}-\theta_{\alpha m})
\int_0^t B^{(m)}_\alpha(\tau)\,d\tau
\right\rbrace
\bigg\rangle_m
\nonumber
\\
&&=
\exp\Big[
-\sum_{\alpha\beta}
(\theta_{\alpha n}-\theta_{\alpha m})
(\theta_{\beta n}-\theta_{\beta m})
g_{\alpha\beta}(t)
\Big],
\end{eqnarray}
thus proving Eq.~(\ref{EqIntegrand}). We have used that 
$$\langle B_\alpha^{(m)}(t)B_\beta^{(m)}\rangle_m=\langle B_\alpha(t)B_\beta\rangle,$$ 
where $\langle\,\cdot\,\rangle$ denotes the average with respect to $\rho_B$ (see Sec.~\ref{SecModel}), and 
\begin{equation}
B_\alpha(t)=e^{iH_B\tau}B_\alpha e^{-iH_B\tau}.
\end{equation}

Now let us derive formula~(\ref{EqZetaNoneq}). 
Since 
\begin{multline*}
e^{-i(H_B+\sum_\alpha\theta_{\alpha n}B_\alpha)t}
\\=e^{-iH_Bt}\,{\rm T}_+
\exp\left\lbrace
-i\sum_\alpha\theta_{\alpha n}
\int_0^t B_\alpha(\tau)\,d\tau
\right\rbrace
\end{multline*}
and
\begin{multline*}
e^{i(H_B+\sum_\alpha\theta_{\alpha n}B_\alpha)t}
\\={\rm T}_-
\exp\left\lbrace
i\sum_\alpha\theta_{\alpha n}
\int_0^t B_\alpha(\tau)\,d\tau
\right\rbrace
e^{iH_Bt},
\end{multline*}
where
\begin{multline}\label{EqChronExpMinus}
{\rm T}_-\exp\left\lbrace
-i\int_0^t f(\tau)d\tau\right\rbrace\\=
1+\sum_{n=1}^\infty(-i)^n
\int_0^td\tau_1\int_0^{\tau_1}d\tau_2\ldots
\int_0^{\tau_{n-1}}d\tau_n\\
f(\tau_n)\cdots f(\tau_2)f(\tau_1), 
\end{multline}
we can express $\zeta_{mnl}(t,\tau)$ as
\begin{equation*}
\begin{split}
\zeta_{mnl}(t,\tau)
=
\bigg\langle
&{\rm T}_-
\exp\left\lbrace
i\sum_\alpha\theta_{\alpha m}
\int_0^t B_\alpha(s)\,ds
\right\rbrace
\\
\times
&{\rm T}_+
\exp\left\lbrace
-i\sum_\alpha\theta_{\alpha n}
\int_0^t B_\alpha(s)\,ds
\right\rbrace
\\
\times
&{\rm T}_-
\exp\left\lbrace
i\sum_\alpha\theta_{\alpha n}
\int_0^{t-\tau} B_\alpha(s)\,ds
\right\rbrace
\\
\times
&{\rm T}_+
\exp\left\lbrace
-i\sum_\alpha\theta_{\alpha l}
\int_0^{t-\tau} B_\alpha(s)\,ds
\right\rbrace
\bigg\rangle
\end{split}
\end{equation*}
Again, the second-order cumulant expansion gives the exact value of this expectation. It consists of the second-order cumulant expansions of the single chronological exponentials and expectation of the products of different first-order expansion terms. We have
\begin{subequations}
\begin{multline}
\biggl\langle
{\rm T}_+
\exp\left\lbrace
-i\sum_\alpha\theta_{\alpha n}
\int_0^{t} B_\alpha(s)\,ds
\right\rbrace
\biggl\rangle
\\=
\exp\Bigg\{
-\sum_{\alpha,\beta}\theta_{\alpha n}\theta_{\beta n} g_{\alpha\beta}(t)
\Bigg\},
\end{multline}
\begin{multline}
\biggl\langle
{\rm T}_-
\exp\left\lbrace
i\sum_\alpha\theta_{\alpha n}
\int_0^{t} B_\alpha(s)\,ds
\right\rbrace
\biggl\rangle
\\=
\exp\Bigg\{
-\sum_{\alpha,\beta}\theta_{\alpha n}\theta_{\beta n} g^*_{\alpha\beta}(t)
\Bigg\},
\end{multline}
\end{subequations}

Let us consider an expectation of a product of first-order expansion terms:
\begin{equation*}
\begin{split}
\bigg\langle
\int_0^{t_1}B_\alpha(s_1)ds_1
&\int_0^{t_2}B_\beta(s_2)ds_2
\bigg\rangle
\\
=&\int_0^{t_1}ds_1\int_0^{t_2}ds_2\,C_{\alpha\beta}(s_1-s_2)
\end{split}
\end{equation*}
\begin{equation*}
\begin{split}
=
g_{\alpha\beta}(t_1)\,
+
&\int_0^{t_1}ds_1\int_{s_1}^{t_2}ds_2\,
C_{\alpha\beta}(s_1-s_2)
\\
=
g_{\alpha\beta}(t_1)\,
+
&\int_0^{t_1}ds_1\int_{0}^{t_2-s_1}ds_2\,
C_{\alpha\beta}(-s_2)
\\
=
g_{\alpha\beta}(t_1)\,
+
&\int_0^{t_2}ds_1\int_{0}^{t_2-s_1}ds_2\,
C_{\alpha\beta}(-s_2)
\\
+
&\int_{t_2}^{t_1}ds_1\int_{0}^{t_2-s_1}ds_2\,
C_{\alpha\beta}(-s_2).
\end{split}
\end{equation*}
The second and the third terms can be transformed into
\begin{multline*}
\int_0^{t_2}ds_1\int_{0}^{s_1}ds_2\,
C_{\alpha\beta}(-s_2)
\\
=
\int_0^{t_2}ds_1\int_{0}^{s_1}ds_2\,
C^*_{\beta\alpha}(s_2)=g_{\beta\alpha}^*(t_2)
\end{multline*}
and
\begin{equation*}
\begin{split}
&\int_{0}^{t_1-t_2}ds_1\int_{0}^{-s_1}ds_2\,
C_{\alpha\beta}(-s_2)
\\
=&\int_{0}^{t_1-t_2}ds_1\int_{0}^{-s_1}ds_2\,
C_{\alpha\beta}(-s_2)
\\
-&\int_{0}^{t_1-t_2}ds_1\int_{0}^{s_1}ds_2\,
C_{\alpha\beta}(s_2)
=-g_{\alpha\beta}(t_1-t_2).
\end{split}
\end{equation*}
Thus,
\begin{multline}\label{Eqh}
\bigg\langle
\int_0^{t_1}B_\alpha(s_1)ds_1
\int_0^{t_2}B_\beta(s_2)ds_2
\bigg\rangle
=
h_{\alpha\beta}(t_1,t_2)\\
\equiv g_{\alpha\beta}(t_1)-g_{\alpha\beta}(t_1-t_2)+g^*_{\beta\alpha}(t_2).
\end{multline}

We have
\begin{equation*}
\begin{split}
\zeta_{mnl}(t,\tau)\\=
\exp\bigg\{
-&\sum_{\alpha\beta}\theta_{\alpha m}\theta_{\beta m}
g_{\alpha\beta}(t)
\\
-&\sum_{\alpha\beta}\theta_{\alpha n}\theta_{\beta n}
\Big[g^*_{\alpha\beta}(t)-g_{\alpha\beta}(t-\tau)\Big]
\\
-&\sum_{\alpha\beta}\theta_{\alpha l}\theta_{\beta l}
g_{\alpha\beta}(t-\tau)
\\
+&\sum_{\alpha\beta}\theta_{\alpha m}\theta_{\beta n}
\Big[h_{\alpha\beta}(t,t)-h_{\alpha\beta}(t,t-\tau)\Big]
\\
+&\sum_{\alpha\beta}\theta_{\alpha m}\theta_{\beta l}
h_{\alpha\beta}(t,t-\tau)
\\
+&\sum_{\alpha\beta}\theta_{\alpha n}\theta_{\beta n}
h_{\alpha\beta}(t,t-\tau)
\\
+&\sum_{\alpha\beta}\theta_{\alpha n}\theta_{\beta l}
\Big[h_{\alpha\beta}(t,t-\tau)-h_{\alpha\beta}(t-\tau,t-\tau)\Big]
\bigg\},
\end{split}
\end{equation*}
which, after the substitution of expression (\ref{Eqh}) for $h_{\alpha\beta}$, gives Eq.~(\ref{EqZetaNoneq}).


\begin{thebibliography}{99}

\bibitem{KatzKosloff}
G.~Katz and R.~Kosloff,
Quantum thermodynamics in strong coupling: Heat transport and refrigeration,
Entropy {\bf 18}, 186 (2016).

\bibitem{Strasberg}
P.~Strasberg, G.~Schaller, N.~Lambert, and T.~Brandes,
Nonequilibrium thermodynamics in the strong coupling and non-Markovian regime based on a reaction coordinate mapping,
New J. Phys. {\bf 18}, 073007 (2016).


\bibitem{Nazir}
D.~Newman, F.~Mintert, and A.~Nazir,
Performance of a quantum heat engine at strong reservoir coupling,
{\href{https://doi.org/10.1103/PhysRevE.95.032139}{Phys. Rev. E {\bf 95}, 032139 (2017)}}.

\bibitem{Dou}
W.~Dou, M.\,A.~Ochoa, A.~Nitzan, and J.\,E.~Subotnik,
Universal approach to quantum thermodynamics in the strong coupling regime,
{\href{https://doi.org/10.1103/PhysRevB.98.134306}{Phys. Rev. B {\bf 98}, 134306 (2018)}}.


\bibitem{Rivas}
A.~Rivas,
Strong coupling thermodynamics of open quantum systems,
{\href{https://doi.org/10.1103/PhysRevLett.124.160601}{Phys. Rev. Lett. {\bf 124}, 160601 (2020)}}.


\bibitem{Redfield}
A.G.~Redfield, 
The theory of relaxation processes,
{\href{https://doi.org/10.1016/B978-1-4832-3114-3.50007-6}{Adv. Magn. Opt. Reson. \textbf{1}, 1--32 (1965)}}.

\bibitem{Davies}
E.~Davies, Markovian master equations,
{\href{https://doi.org/10.1007/BF01608389}{Commun. Math. Phys. \textbf{39}, 91--110 (1974)}}.

\bibitem{Davies2}
E.~Davies, Markovian master equations. II,
{\href{https://doi.org/10.1007/BF01351898}{Math. Ann. \textbf{219}, 147--158 (1976)}}.


\bibitem{TaniKubo}
Y.~Tanimura and R.~Kubo,
Time evolution of a quantum system in contact with a nearly Gaussian-Markoffian noise bath,
J. Phys. Soc. Jpn. {\bf 58}, 101--114 (1989).

\bibitem{IFl}
A.\,Ishizaki and G.\,R.\,Fleming, Unified treatment of quantum coherent and incoherent hopping dynamics in electronic energy transfer: Reduced hierarchy equation approach, {\href{http://dx.doi.org/10.1063/1.3155372}{J. Chem. Phys. \textbf{130}, 234111 (2009)}}.

\bibitem{Tani}
Y.~Tanimura,
Numerically ``exact'' approach to open quantum dynamics: The hierarchical equations of motion (HEOM),
J. Chem. Phys. {\bf 153}, 020901 (2020).

\bibitem{Vega}
I.\,de Vega and D.~Alonso,
Dynamics of non-Markovian open quantum systems,
{\href{https://doi.org/10.1103/RevModPhys.89.15001}{Rev. Mod. Phys. {\bf 89}, 15001 (2017)}}.


\bibitem{Tamapre}
D.~Tamascelli, A.~Smirne, J.~Lim, S.\,F.~Huelga, and M.\,B.~Plenio,
Efficient simulation of finite-temperature open quantum systems,
{\href{https://doi.org/10.1103/PhysRevLett.123.090402}{Phys. Rev. Lett. {\bf 123}, 090402 (2019)}}.


\bibitem{Tama}
F.~Mascherpa, A.~Smirne, A.\,D.~Somoza, P.~Fern\'{a}ndez-Acebal, 
S.~Donadi, D.~Tamascelli, S.\,F.~Huelga, and M.\,B.~Plenio,
Optimized auxiliary oscillators for the simulation of general open quantum systems,
{\href{https://doi.org/10.1103/PhysRevA.101.052108}{Phys. Rev. A {\bf 101}, 052108 (2020)}}.


\bibitem{GarrawayPetruc}
G.~Pleasance, B.\,M.~Garraway, and F.~Petruccione,
Generalized theory of pseudomodes for exact descriptions of non-Markovian quantum processes,
{\href{https://doi.org/10.1103/PhysRevRes.2.043058}{Phys. Rev. Res. {\bf 2}, 043058 (2020)}}.


\bibitem{TereFinT}
A.\,E.~Teretenkov, 
Integral representation of finite temperature non-Markovian evolution of some systems in rotating wave approximation,
Lobachevskii J. Math. {\bf 41}, 2397--2404 (2020).

\bibitem{TereSeveralBath}
A.\,E.~Teretenkov, 
Exact non-Markovian evolution with several reservoirs,
Phys. Part. Nucl. {\bf 51}, 479--484 (2020).

\bibitem{Lambert}
J.~Iles-Smith, N.~Lambert, and A.~Nazir,
Environmental dynamics, correlations, and the emergence of noncanonical equilibrium states in open quantum systems,
{\href{https://doi.org/10.1103/PhysRevA.90.032114}{Phys. Rev. A {\bf 90}, 032114 (2014)}}.


\bibitem{Polaron0}
S.~Jang, Y.-C.~Cheng, D.\,R.~Reichman, and J.\,D.~Eaves, 
Theory of coherent resonance energy transfer,
J. Chem. Phys. {\bf 129}, 101104 (2008).

\bibitem{Polaron}
A.~Kolli, A.~Nazir, and A.~Olaya-Castro, 
Electronic excitation dynamics in multichromophoric systems described via a polaron-representation master equation,
J. Chem. Phys. {\bf 135}, 154112 (2011).

\bibitem{Gorini}
V.\,Gorini, A.\,Frigerio, M.\,Verri, A.\,Kossakowski, and E.\,C.\,G.\,Sudarshan, 
Properties of quantum Markovian master equations,
{\href{https://doi.org/10.1016/0034-4877(78)90050-2}{Rep. Math. Phys. \textbf{13}, 149--173 (1978)}}.


\bibitem{Palmer}
P.\,F.\,Palmer,
The singular coupling and weak coupling limits,
{\href{https://doi.org/10.1063/1.523296}{J. Math. Phys. \textbf{18}, 527--529 (1977)}}.

\bibitem{AccFriLu}
L.\,Accardi, A.\,Frigerio, and Y.\,G.\,Lu,
On the relation between the singular and the weak coupling limits,
{\href{https://doi.org/10.1007/BF00047202}{Acta Appl. Math. \textbf{26}, 197--208 (1992)}}.

\bibitem{Dumcke}
R.\,D\"{u}mcke, The low-density limit for an $N$-level system interacting with a free Bose or Fermi gas, Commun. Math. Phys. {\bf 97}, 331--359 (1985).

\bibitem{Pechen}
L.\,Accardi, A.\,N. Pechen, and I.\,V.\,Volovich, Quantum stochastic equation for the low density limit, J. Phys. A {\bf 35}, 4889--4902 (2002).

\bibitem{BP}
H.-P.~Breuer and F.~Petruccione,
\emph{The Theory of Open Quantum Systems}
(Oxford University Press, 2002).

\bibitem{LambertNori}
N.~Lambert, S.~Ahmed, M.~Cirio, and F.~Nori, 
Modelling the ultra-strongly coupled spin-boson model with unphysical modes,
\href{https://doi.org/10.1038/s41467-019-11656-1}{Nature Comm. {\bf 10}, 3721 (2019)}.

\bibitem{Kawai}
K.~Goyal and R.~Kawai, Steady state thermodynamics of two qubits strongly coupled to bosonic environments, \href{https://dx.doi.org/10.1103/PhysRevResearch.1.033018}{Phys. Rev. Res. {\bf 1}, 033018 (2019)}.

\bibitem{CresserAnders}
J.\,D.~Cresser and J.~Anders, Weak and ultrastrong coupling limits of the quantum mean force Gibbs state, 
{\href{https://arxiv.org/abs/2104.12606}{arXiv:2104.12606}}.

\bibitem{UstrongChargedQubits}
J.\,Yu, F.\,A.\,C\'{a}rdenas-L\'{o}pez, C.\,K.\,Andersen, E.\,Solano, A.\,Parra-Rodriguez, Charge qubits in the ultrastrong coupling regime, 
\href{https://arxiv.org/abs/2105.06851}{arXiv:2105.06851}


\bibitem{MK}
V.~May and O.~K{\"u}hn, Charge and Energy Transfer Dynamics in Molecular Systems (Wiley-VCH, Weinheim, 2011).


\bibitem{Valkunas}
L.~Valkunas, D.~Abramavicius, and T.~Man{\v c}al,
Dynamical Excitation Dynamics and Relaxation (Wiley-VCH Verlag, Berlin, 2013).

\bibitem{YangFl}
M.~Yang and G.R.~Fleming, 
Influence of photons on exciton transfer dynamics: comparison of the Redfield, F{\"o}rster, and modified Redfield equations,
Chem. Phys. {\bf 275}, 355--372 (2002).


\bibitem{NovoGrond}
V.\,I.~Novoderezhkin and R.~van Grondelle, 
Physical origins and models of energy transfer in photosynthetic light-harvesting,
Phys. Chem. Chem. Phys. {\bf 12}, 7352--7365 (2010).


\bibitem{Seibt}
J.~Seibt and T.~Man{\v c}al, 
Ultrafast energy transfer with competing channels: Non-equilibrium F{\"o}rster and modified Redfield theories,
J. Chem. Phys. {\bf 146}, 174109 (2017).


\bibitem{Forster1}
Th. F{\"o}rster, 
Energiewanderung und Fluoreszenz,
Naturwissenschaften {\bf 33} (6), 166--175 (1946).

\bibitem{Forster2}
Th. F{\"o}rster, 
Zwischenmolekulare Energiewanderung und Fluoreszenz,
Ann. Phys. {\bf 437} (1), 55--75 (1948).


\bibitem{mRedf}
W.M.~Zhang, T.~Meier, V.~Chernyak, and S.~Mukamel,
Exciton-migration and three-pulse femtosecond optical spectroscopy of photosynthetic antenna complexes,
J. Chem. Phys. {\bf 108}, 7763--7774 (1998).



\bibitem{NovoGrond2013}
V.\,I.~Novoderezhkin and R.~van Grondelle, 
Spectra and Dynamics in the B800 Antenna: Comparing Hierarchical Equations, {Redfield} and {F{\"o}rster} Theories,
J. Phys. Chem. B {\bf 117} (38), 11076--11090 (2013).


\bibitem{NovoGrond2017}
V.\,I.~Novoderezhkin and R.~van Grondelle, 
Modeling of excitation dynamics in photosynthetic light-harvesting complexes: exact versus perturbative approaches,
J. Phys. B {\bf 50}, 124003 (2017).


\bibitem{TrushJCP}
A.~Trushechkin,
Calculation of coherences in F\"orster and modified Redfield theories of excitation energy transfer,
J. Chem. Phys. {\bf 151}, 074101 (2019).


\bibitem{Decoherence}
M.~Schlosshauer, Quantum decoherence, \href{https://dx.doi.org/10.1016/j.physrep.2019.10.001}{Phys. Rep. {\bf 831}, 1--57 (2019)}.


\bibitem{TDdecoh}
M.\,Popovic, M.\,T.\,Mitchison, and J.\,Goold, Thermodynamics of decoherence, {\href{https://arxiv.org/abs/2107.14216}{arXiv:2107.14216}}.

\bibitem{ForsterNoneq}
S. Jang, Y. J. Jung, and R. J. Silbey, 
Nonequilibrium generalization of F{\"o}rster--Dexter theory for excitation energy transfer, Chem. Phys. {\bf 275}, 319--332 (2002).



\bibitem{RH}
A.~Rivas and S.\,F.~Huelga, Open Quantum Systems: An Introduction (Springer, 2012).

\bibitem{Bach}
V.\,Bach, J.\,Fr\"ohlich, and I.\,M.\,Sigal, Return to equilibrium,
{\href{https://doi.org/10.1063/1.533334}{J. Math. Phys. \textbf{41}, 3985--4060 (2000)}}.

\bibitem{Frohlich}
J.\,Fr\"ohlich and M.\,Merkli, Another return of ``Return to equilibrium'',
{\href{https://doi.org/10.1007/s00220-004-1176-6}{Commun. Math. Phys. \textbf{251}, 235--262 (2004)}}.


\bibitem{EspositoGaspard}
M.\,Esposito and P.\,Gaspard
Quantum master equation for a system influencing its environment,
{\href{https://doi.org/10.1103/PhysRevE.68.066112}{Phys. Rev. E {\bf 68}, 066112 (2003)}}.


\bibitem{Budini}
A.\,A.\,Budini,
Random Lindblad equations from complex environments,
{\href{https://doi.org/10.1103/PhysRevE.72.056106}{Phys. Rev. E {\bf 72}, 056106 (2005)}}.


\bibitem{Breuer2006}
H.-P.\,Breuer, J.\,Gemmer, and M.\,Michel
Non-Markovian quantum dynamics: Correlated projection superoperators and Hilbert space averaging,
{\href{https://doi.org/10.1103/PhysRevE.73.016139}{Phys. Rev. E {\bf 73}, 016139 (2006)}}.


\bibitem{Breuer2007}
H.-P.\,Breuer,
Non-Markovian generalization of the Lindblad theory of open quantum systems,
{\href{https://doi.org/10.1103/PhysRevA.75.022103}{Phys. Rev. A {\bf 75}, 022103 (2007)}}.

\bibitem{StrasbergHier}
A.\,Riera-Campeny, A.\,Sanpera, and P.\,Strasberg,
``Open quantum systems coupled to finite baths: A hierarchy of master equations'',
{\href{https://arxiv.org/abs/2108.01890}{arXiv:2108.01890}}.



\bibitem{MerkliNesterovDimer}
M.\,Merkli, G.\,P.\,Berman, R.\,T.\,Sayre, S.\,Gnanakaran, M.\,K\"{o}nenberg, A.\,I.\,Nesterov, H.\,Song,
Dynamics of a chlorophyll dimer in collective and local thermal environments,
J. Math. Chem. {\bf 54} 866--917 (2016).

\bibitem{MerkliDimer}
M.\,Merkli, Effective evolution of open dimers,
Contemp. Math. \textbf{717}, 323--338 (2018).




\bibitem{QEffBio}
R.\,Mohseni, Y.\,Omar, G.\,Engel, and M.\,B.\,Plenio (Eds.),
Quantum Effects in Biological Systems (Cambridge University Press, 2014).

\bibitem{Banchi}
L.\,Banchi, G.\,Costagliola, A.\,Ishizaki, and P.\,Giorda, 
An analytical continuation approach for evaluating emission lineshapes of molecular aggregates and the adequacy of multichromophoric {F{\"o}rster} theory, J. Chem. Phys. {\bf 138}, 184107 (2013).

\bibitem{Renger2008}
G.\,Raszewski and T.\,Renger, 
Light Harvesting in Photosystem II Core Complexes Is Limited by the Transfer to the Trap:? Can the Core Complex Turn into a Photoprotective Mode? J. Am. Chem. Soc. {\bf 130}, 4431--4446 (2008).

\bibitem{Jang2014}
S.\,Jang, S.\,Hoyer, G.\,Fleming, and K.\,B.\,Whaley,
Generalized master equation with non-Markovian multichromophoric F\"orster resonance energy transfer for modular exciton densities,
{\href{https://doi.org/10.1103/PhysRevLett.113.188102}{Phys. Rev. Lett. {\bf 113}, 188102 (2014)}}.


\bibitem{Yang2003}
M.\,Yang, A.Damjanoci\'c, H.\,M.\,Vaswani, and G.\,R.\,Fleming, Biophys. J. {\bf 85}, 140--158 (2003).

\bibitem{Renger2011}
T.\,Renger, M.\,Madjet, A.\,Knorr, and F.\,M\"uh, How the molecular structure determines the flow of excitation energy in plant light-harvesting complex II,  J. Plant Physiol. {\bf 168}, 1497--1509
(2011).


\bibitem{Uni}
A.\,Trushechkin, Unified Gorini-Kossakowki-Lindblad-Sudarshan quantum master equation beyond the secular approximation, {\href{https://doi.org/10.1103/PhysRevA.103.062226}{Phys. Rev. A {\bf 113}, 062226 (2021)}}.

\bibitem{DannMegierKosloff}
R.\,Dann, N.\,Megier, and R.\,Kosloff, 
Non-Markovian dynamics under time-translation symmetry,
{\href{https://arxiv.org/abs/2106.05295}{arXiv:2106.05295}}.

\bibitem{Mukamel}
S.\,Mukamel, Principles of Nonlinear Optical Spectroscopy (Oxford University Press, New York, 1995).

\end{thebibliography}
\end{document}